\documentclass[10pt]{article}
\usepackage[utf8]{inputenc}
\usepackage{graphicx}
\usepackage[utf8]{inputenc}
\usepackage{amsmath}
\usepackage{amsfonts}
\usepackage{amssymb}
\usepackage{graphicx}
\usepackage{amsmath, physics, siunitx}
\usepackage{epstopdf}
\AppendGraphicsExtensions{.tif}
\usepackage{setspace}
\usepackage{authblk}
\usepackage{caption}
\usepackage{float}
\usepackage{cite}
\usepackage[left=2.0cm,right=2.0cm,top=2cm,bottom=2cm]{geometry}
\usepackage{color}
\usepackage{color,soul}
\usepackage{multicol}
\usepackage{subcaption}

\title{\textbf{Coupled Dipole Oscillation Enabled Scintillation Compensation in Turbulence-Impacted Laser Beam}}

\author[1,*]{Shouvik Sadhukhan}

\author[2]{C. S. Narayanamurthy}

\affil[1, 2]{\small{Applied and Adaptive Optics Laboratory, Department of Physics, Indian Institute of Space Science and Technology (IIST), P.O: Valiamala, Trivandrum - 695547, State: Kerala; India}}
\affil[1]{\small{Email: shouvikphysics1996@gmail.com}}
\affil[2]{\small{Email: naamu.s@gmail.com}}
\affil[*]{\small{Corresponding Author Email: shouvikphysics1996@gmail.com}}

\begin{document}
\maketitle

\begin{abstract}
The present work develops a rigorous framework for collective dipole oscillations in dense dielectric media by extending the Lorentz oscillator model to include both quadratic (second-order) and cubic (third-order) nonlinear restoring forces. Dipole-dipole interactions are incorporated via the dyadic Green’s function, leading to a nonlocal description of the medium. Starting from the generalized oscillator equation, the system is expressed in matrix form with an effective stiffness matrix. Using a harmonic steady-state ansatz, nonlinear terms are approximated in the frequency domain. Diagonalization provides collective normal modes, which serve as a basis for perturbative expansion: zeroth-order captures the linear response, while higher orders describe nonlinear corrections. The polarization is expressed through corrected mode amplitudes, yielding an effective nonlinear nonlocal susceptibility kernel. Finally, the scalar Green’s function formalism is applied to derive the structured output field. This approach bridges microscopic dipole dynamics with macroscopic optical propagation, offering key insights into nonlinear light–matter interaction and turbulence-influenced field evolution.\\

\end{abstract}

\section{Introduction}
Optical communication systems operating in real-world environments often face severe performance degradation due to atmospheric turbulence, which causes random fluctuations in the intensity, phase, and polarization of propagating beams. This results in beam wander, scintillation, and loss of coherence, ultimately reducing the reliability of free-space optical links. The work presented in this study introduces a rigorous theoretical and experimental framework for mitigating such turbulence-induced distortions by leveraging collective dipole oscillations in dense dielectric media.\cite{6,7,8,9,10,11,12,13,14,15,16,17,18,19,20}\par

This work presents a comprehensive range of studies that collectively advance the fundamental understanding of optical forces, electromagnetic dipoles, and wave-matter interactions. Beginning with mathematical and statistical frameworks, Vishwakarma and Moosath propose innovative methods for quantifying distances in the space of Gaussian mixtures, pivotal for signal analysis and probability theory \cite{1}. Early studies by Korotkova and collaborators systematically examine the evolution of intensity fluctuations and Stokes parameter statistics of random and quasi-monochromatic electromagnetic beams in turbulent or complex media, providing extensive insights on coherence, polarization, and the role of environmental randomness \cite{2,3,4,5,6,7,8,9,10}. The efforts to forge new calculi for optical angular momentum transformations via Jones and Stokes-Mueller matrices have also expanded the mathematical toolkit for polarized beam diagnostics \cite{11,12,13,14,15}. Abbasirad et al.'s investigation into dipole emission near dielectric metasurfaces with scanning near-field optical microscopy underscores advanced experimental techniques for probing electromagnetic responses at the nanoscale \cite{16}, while Scheel and Buhmann's treatise on macroscopic quantum electrodynamics broadens the conceptual base for light-matter coupling in engineered and natural settings \cite{17}. Classic works such as Levine and Schwinger's aperture diffraction theory and Sipe's macroscopic electromagnetic description of resonant dielectrics have shaped the theoretical foundations that underpin contemporary treatments of optical scattering and absorption phenomena \cite{18,19,20,21,22,23,24,25,26,27,28,29,30}.\par

Multiple contributions address the dynamics, enhancement, and nonlinearities of dipole-dipole interactions as well as the ramifications for metamaterials and nanocomposites. Poddubny et al. propose microscopic models for Purcell enhancement within hyperbolic metamaterials, crucial for light emission control \cite{20}. Martin and Piller, as well as Paulus et al., deliver accurate computational techniques for electromagnetic scattering and Green’s tensor evaluation in complex stratified environments \cite{21,22}. The evolving understanding of the Lorentz oscillator, including nonlinear effects and the ramifications of radiative damping, is enriched by Smith and Schelew et al. \cite{23,29}. Studies on optical forces—including fundamental reviews, force and torque predictions by spinning fields, and experimental observations of light-induced dipole forces in ultracold gases—are presented by Tamura \cite{33}, Lembessis \cite{34}, Saif \cite{35}, Matsumori \cite{37}, Canaguier-Durand \cite{38}, Kumar \cite{39}, and Maiwoger \cite{40}, demonstrating the breadth and technological relevance of optomechanical manipulation. Further, several references deepen the discourse on discrete dipole approximations, time-varying polarizabilities, quadrupolar responses, and nonlinearity in nanoantennas and metasurfaces, such as Bowen \cite{25}, Mirmoosa \cite{26}, Achouri \cite{30}, and Sain \cite{45}. Collectively, these papers establish a rigorous, multidisciplinary framework that informs the present investigation and inspires future advances in optical physics, material science, and wave-based technologies.\cite{31,32,33,34,35,36,37,38,39,40,41,42,43,44,45}\par

The theoretical foundation begins with the Lorentz oscillator model, which describes the oscillation of bound electrons under an external electromagnetic field. While the classical model accounts for linear restoring forces, real molecular systems—such as those in polymer media—also exhibit nonlinear restoring contributions. To address this, the model is extended to include both quadratic (second-order) and cubic (third-order) nonlinearities. These anharmonic corrections capture more realistic electron cloud dynamics and their nonlinear interaction with incident fields. A central theme of the work is the role of dipole–dipole interactions, which are introduced through the dyadic Green’s function formalism. Unlike the scalar Green’s function, the dyadic representation preserves the vector nature of electromagnetic fields and accurately accounts for near-field, induction-zone, and far-field coupling between oscillating dipoles. This nonlocal description of the medium enables the derivation of collective oscillation modes, obtained by diagonalizing the effective stiffness matrix of the system. The resulting eigenmodes form a natural basis for perturbative analysis: zeroth-order solutions describe the linear response, while higher orders incorporate nonlinear corrections that manifest in effective nonlocal susceptibility kernels. The output field is reconstructed using the scalar Green’s function for free-space propagation, linking the microscopic dipole oscillations to macroscopic beam statistics. This theoretical framework predicts that when light propagates through a dielectric medium—such as PMMA (poly(methyl methacrylate)) rods—the collective synchronization of dipole oscillations can significantly reduce scintillation caused by turbulence. Physically, synchronized dipoles impose inertia on fluctuating external fields, slowing down centroid shifts and suppressing intensity variance.\cite{46,47,48,49,50,51,52,53,54,55}\par

When an electromagnetic field propagates through a transparent or semi-transparent medium, its electric field interacts with the molecular structures of the material. This interaction perturbs the electron clouds, breaking the stationary symmetries of the molecules and giving rise to induced dipoles. As the external field oscillates, it drives corresponding dipole oscillations within the medium. Since the molecules are not isolated but interconnected, they are able to exchange energy, leading to the emergence of a coupled dipole oscillation model. In such a system, the coupling imposes constraints on the motion of individual dipoles, effectively altering the system’s degrees of freedom. The coupling coefficients introduce non-diagonal terms into the governing equations, meaning the standard Cartesian coordinates can no longer serve as an orthonormal basis for describing the dynamics. Consequently, the system cannot be represented by independent dynamical equations without further transformation. To resolve this, a process of diagonalization is applied to the coupled equations, redefining the degrees of freedom and establishing a new orthonormal reference frame in which the system can be analyzed more clearly. The diagonalized representation reveals a set of collective oscillatory modes that describe the synchronized dipole behavior. Unlike the original random dipole oscillations, which are chaotic and uncorrelated, the coupled mode dynamics reflect coherent, collective oscillations stabilized by the mutual interactions among dipoles. This transformation from random to coupled modes highlights the importance of diagonalization in simplifying the description of complex dipole systems, allowing the emergent ordered dynamics to be distinguished from the initial disorder induced by the external field. Ultimately, this framework provides a powerful way to understand how microscopic molecular interactions shape the macroscopic electromagnetic response of materials.\cite{56,57,58,59,60,61,62,63,64,65,66,67,68,69,70,71,72,73,74,75} \par

The present work demonstrated as in section \ref{2} detail theoretical discussion of the work have been given. The statistical background has been discussed in section \ref{3}. The section \ref{4} contains experimental details where the results analysis have been added in section \ref{5}. Finally, the paper is concluded into section \ref{6}.

\section{Theoretical Background}\label{2}
The fundamental theoretical work begings with the dynamical equation of the Lorentz Oscillation model under external field as follows.

\begin{equation}
    m \ddot{\mathbf{r}}_i + m \gamma \dot{\mathbf{r}}_i + m \omega_0^2 \mathbf{r}_i = -e \mathbf{E}_{\text{ext}}(\mathbf{r}_i, t)
\end{equation}

Here $\vec{\textbf{r}}$ is the relative position vector with respect to the nucleus, $m\rightarrow$ effective electron mass where, $\frac{1}{m}=\frac{1}{\hbar^2}\frac{\partial^2 E(\kappa)}{\partial\kappa^2}$ with $E(\kappa)$ is the electron dispersion relation which is generally defined by the curvature of the electronic band structure, $\omega_0 \rightarrow$ natural (resonant) frequency of bound electron oscillation and $\gamma\rightarrow$ damping constant (collisions, radiation loss). The term which is contained with natural frequency represents the restoring effect on the external field which causes the perturbation on the electron cloud to induce dipole moment.

\subsection{Lorentz Anharmonic Dipole Oscillator}
In general, not all materials and their molecular structures can produce just linear restoring forces on their bound electron clouds due to propagation of electromagnetic wave through them. The complex molecular structures can produce higher order restoring effects too. Hence, in the present context, we introduce second order and third order nonlinearities in restoring forces for the molecules of the PMMA (Poly(methyl methacrylate)) rod. Thus, the Generalized Forced Anharmonic Oscillator differential equation can be written as follows.

\begin{equation}
    m \ddot{\mathbf{r}}_i + m \gamma \dot{\mathbf{r}}_i + m \omega_0^2 \mathbf{r}_i + \beta_i |\mathbf{r}_i| \mathbf{r}_i + \alpha_i |\mathbf{r}_i|^2 \mathbf{r}_i = -e \mathbf{E}_{\text{ext}}(\mathbf{r}_i, t)
\end{equation}
Here, we have,
\begin{itemize}
    \item $\beta_i\rightarrow$ second-order nonlinearity coefficient.
    \item $\alpha_i\rightarrow$ third-order nonlinearity coefficient.
\end{itemize}

Whenever an electromagnetic field enters into a transparent or semi-transparent medium, the electric field of the EM wave interacts with the molecules of the medium. The interaction between the electric field and the material/medium molecules produces an electric field perturbation on the electron clouds. The presence of perturbations breaks the stationary symmetries of the molecules and the corresponding electron clouds. Thus, the breaking of symmetries generates dipoles inside the molecules. The oscillatory amplitude variation of the external field provides dipole oscillations inside the molecules of the propagating medium. The molecules inside the material are inter-connected and hence, they can exchange energy. Thus, coupled dipole oscillation model comes into scenario. 

\subsection{Lorentz Coupled Anharmonicity}
The electric field propagation through a medium in presence of dipole-dipole interaction and coupling is equivalent phenomena as the electric field propagation through a medium under presence of initial charge densities. Thus, the field propagation through such medium must be discussed with the Vector Helmholtz equation. In this context of electromagnetic wave propagation through medium, the electromagnetic field spatial distribution can be analytically discussed using following parent differential equations.

\begin{equation}
    \nabla\times\nabla\times\mathbf{E(r)}-\kappa^2\mathbf{E}=i\omega\mu\mathbf{J(r)}
\end{equation}

In our present discussion, we are dealing with the dipole-dipole coupling which produces additional forces on localized dipoles. Hence, for localized reference frame we need to follow Vector Helmholtz Equation to study the electromagnetic field propagation through the medium to include localized dipole-dipole interaction. The presence of localized dipoles on a particular position must be effected by the presence of dipole-dipole coupling or interaction with other localized dipoles on other sights. Thus, the force field from surrounding dipoles on a particular dipole should be discussed using Dyadic Green's Function $\mathbf{G}(\mathbf{r}_i, \mathbf{r}_j)$. Such Green's function is responsible for optical field propagation in presence of initial charge density through Vector Helmholtz equation. From the vector helmholtz equation we can write following differential equation for green's function.

\begin{equation}
    \left[ \nabla \times \nabla \times - k^2 \mathbf{I} \right] 
\overline{\overline{G}}(\mathbf{r},\mathbf{r}') 
= \mathbf{I} \delta(\mathbf{r}-\mathbf{r}')
\end{equation}

Where the source should be represented as follows.

\begin{equation}
    \mathbf{E}(\mathbf{r}) = i\omega \mu_0 \int 
\overline{\overline{G}}(\mathbf{r},\mathbf{r}') \cdot \mathbf{J}(\mathbf{r}') 
\, d^3\mathbf{r}'
\end{equation}

$\overline{\overline{G}}$ generalizes the scalar Green’s function to vector fields, ensuring transversality and coupling between components. The solution of the green's differential equation can be derived as follows which basically produces the expanded scalar green's function $g(\mathbf{r},\mathbf{r}') = \frac{e^{ik|\mathbf{r}-\mathbf{r}'|}}{4\pi |\mathbf{r}-\mathbf{r}'|}$.

\begin{equation}
    \overline{\overline{G}}(\mathbf{r},\mathbf{r}') 
= \left( \mathbf{I} + \frac{1}{k^2} \nabla \nabla \right) \frac{e^{ik|\mathbf{r}-\mathbf{r}'|}}{4\pi |\mathbf{r}-\mathbf{r}'|}
\end{equation}

Here, we have,
\begin{itemize}
    \item $\mathbf{I} g$: Isotropic spherical wave contribution (free-space spreading).
    \item $\frac{1}{k^2}\nabla\nabla g$: Longitudinal correction ensuring $\nabla \cdot \mathbf{E} = 0$ (no free charges in homogeneous medium).
    \item Together: Maintain vector nature of electromagnetic field.
\end{itemize}

This function maps a point current source $\mathbf{J}(\mathbf{r}')$ to the resulting vector field $\mathbf{E}(\mathbf{r})$ including polarization, coupling, and radiation effects. The General Expansion of dyadic green's function can be written as follows with $r = |\mathbf{r}-\mathbf{r}'|$, $\hat{\mathbf{r}} = \frac{\mathbf{r}-\mathbf{r}'}{r}$.

\begin{equation}
    \overline{\overline{G}}(\mathbf{r},\mathbf{r}') 
= \frac{e^{ikr}}{4\pi r}
\left[
\left( \mathbf{I} - \hat{\mathbf{r}}\hat{\mathbf{r}} \right)
\left(1 + \frac{i}{kr} - \frac{1}{(kr)^2}\right)
+ \hat{\mathbf{r}}\hat{\mathbf{r}}
\left(1 + \frac{3i}{kr} - \frac{3}{(kr)^2}\right)
\right]
\end{equation}

From this expression we can get,
\begin{itemize}
    \item \textbf{Near Field (Quasi-static, $kr \ll 1$):} Electrostatic dipole-like behavior as \[
    \overline{\overline{G}} \sim \frac{1}{4\pi r^3}(3\hat{\mathbf{r}}\hat{\mathbf{r}} - \mathbf{I})
    \]
    \item \textbf{Induction Zone (Intermediate, $kr \sim 1$):} Magnetic induction and reactive energy storage as \[
    \overline{\overline{G}} \sim \frac{1}{4\pi r^2}(ik)(3\hat{\mathbf{r}}\hat{\mathbf{r}} - \mathbf{I})
    \]
    \item \textbf{Far Field (Radiation Zone, $kr \gg 1$):} Transverse spherical wave radiation as \[
    \overline{\overline{G}} \sim \frac{e^{ikr}}{4\pi r}
    \left( \mathbf{I} - \hat{\mathbf{r}}\hat{\mathbf{r}} \right)
    \]
\end{itemize}
The displacement of the $i$-th electron from equilibrium produces a dipole moment should be considered as $\mathbf p_i(\omega) = -e \mathbf r_i(\omega)$ for arbitrary angular frequency $\omega$. Hence, point dipole at $\mathbf r_j$ corresponds to a current density with frequency $\omega$ is as $\mathbf J(\mathbf r, t) = \frac{\partial \mathbf P (r,t)}{\partial t}$ with $\mathbf P (r, t) = \mathbf p_j(r,t) \, \delta(\mathbf r - \mathbf r_j)$. As dipole is oscillation with the same frequency as of input, we can get the following current density.

\begin{equation}
\mathbf J(\mathbf r,\omega) = -i\omega \mathbf p_j \, \delta(\mathbf r - \mathbf r_j).
\end{equation}

Substituting this in green's function based propagator equation of electric field gives,

\begin{equation}
\mathbf E(\mathbf r) 
= \mu_0 \omega^2 \, \mathbf G(\mathbf r,\mathbf r_j)\cdot \mathbf p_j(\mathbf r) = \mathbf E^{(j)}_{sc} (\mathbf r_j)
\end{equation}

Thus, the field produced at $\mathbf r$ by dipole $j$ is directly proportional to its dipole moment. The total field at the position of the $i$-th dipole is the sum of the external incident field and the scattered fields due to all other dipoles is as $\mathbf E_{\text{tot}}(\mathbf r_i,\omega) 
= \mathbf E_{\text{ext}}(\mathbf r_i,\omega)
+ \sum_{j\neq i}\mathbf E_{\text{sc}}^{(j)}(\mathbf r_i,\omega)$. Since $\mathbf p_j = -e \mathbf r_j$, hence, we have $\mathbf E_{\text{sc}}^{(j)}(\mathbf r_i,\omega) 
= - e \, \mu_0 \omega^2 \, \mathbf G(\mathbf r_i,\mathbf r_j)\cdot \mathbf r_j$. Therefore, we get,

\begin{equation}
\mathbf E_{\text{tot}}(\mathbf r_i,\omega) 
= \mathbf E_{\text{ext}}(\mathbf r_i,\omega) 
- e \, \mu_0 \omega^2 \sum_{j\neq i}\mathbf G(\mathbf r_i,\mathbf r_j)\cdot \mathbf r_j
\end{equation}

The force acting on the $i$-th electron is $\mathbf F_i(\omega) = (-e)\mathbf E_{\text{tot}}(\mathbf r_i,\omega)$. Now, substituting the expression for $\mathbf E_{\text{tot}}$ gives final expression of force as,

\begin{align}
\mathbf F_i(\omega) 
&= (-e)\mathbf E_{\text{ext}}(\mathbf r_i,\omega)
+ e^2 \mu_0 \omega^2 \sum_{j\neq i}\mathbf G(\mathbf r_i,\mathbf r_j)\cdot \mathbf r_j. 
\end{align}

As in the present framework, we are propagating a light beam through a medium. After the entrance of that beam, it generates dipoles inside the material. The dipole-dipole coupling causes the intra-medium electromagnetic field propagation. On the other hand, the dipole-dipole coupling influences the resultant oscillation modes of the dipoles when a randomly polarized light generates randomly distributed dipoles inside the medium. Thus, the final differential equation of Generalized Anharmonic Dipole Oscillator with Dipole-Dipole Coupling can be written as follows with dipole-dipole coupling coefficient as Dyadic Greens Function $\mathbf{G}(\mathbf{r}_i, \mathbf{r}_j)$.

\begin{equation}
    m \ddot{\mathbf{r}}_i + m \gamma \dot{\mathbf{r}}_i + m \omega_0^2 \mathbf{r}_i + \beta_i |\mathbf{r}_i| \mathbf{r}_i + \alpha_i |\mathbf{r}_i|^2 \mathbf{r}_i = -e \mathbf{E}_{\text{ext}}(\mathbf{r}_i, t) + e^2 \mu_0 \omega^2 \sum_{j \ne i} \mathbf{G}(\mathbf{r}_i, \mathbf{r}_j) \cdot \mathbf{r}_j(t)
\end{equation}

\subsection{Gradient Force Impact on Dipole Moment}
The gradient of the electric field with non-uniform amplitude produces an additional impact on the coupling between the nearest neighbour dipoles. The presence of non-zero gradients and higher-orders derivatives of the electric field produces non-uniform dipole moments on spatially distributed dipoles. These non-uniformities modify the conventional dipole-dipole coupling by introducing gradient-based repelling forces. Hence, the gradient force contribution should be considered into the generalized Lorentz force dynamics. In the present context of the formalism for the Generalized Anharmonic Dipole Oscillator with Dipole-Dipole Coupling we conventionally found two forces that are modulating the localized dipole moments. Those forces are as follows.
\begin{itemize}
    \item $\mathbf F_{Ext}=-e \mathbf{E}_{\text{ext}}(\mathbf{r}_i, t)$ which is the externally applied field due to propagation of light fields.
    \item $\mathbf F_{Coupling}= e^2 \mu_0 \omega^2 \sum_{j \ne i} \mathbf{G}(\mathbf{r}_i, \mathbf{r}_j) \cdot \mathbf{r}_j(t)$ which is the dipole-dipole coupled forces that are applied from surrounding localized dipoles on the specific dipole.
\end{itemize}
Now, if the external force field i.e., the propagating field spatial distribution is non-uniform or variable, the external force field can be expanded into corresponding Taylor's expansion. The higher order derivatives of the spatially varying external field will contribute additional force on coupling. That forces act as Gradient Coupling Force. Hence, total force on a point $r_i$ can be written as follows ($HO=$ Higher orders).

\begin{equation}
    \mathbf F_{TotExt}=-e \mathbf{E}_{\text{ext}}(\mathbf{r}_{i}, t)-\sum_{j \ne i}\left [(e(\vec{\mathbf r}_i -  \vec{\mathbf  r}_j)\cdot\nabla)\mathbf{E}_{\text{ext}}(\mathbf{r}_{j}, t)+ HO\right]=\mathbf F_{Ext}+\mathbf F_{HO}
\end{equation}

Thus, in the present context, the total coupling forces can be written as a summation of Gradient Coupling Force and Dipole-Dipole Coupling Forces as follows.

\begin{equation}
    \mathbf F_{TotCoupling} = \mathbf F_{Coupling} + \mathbf F_{HO} = e^2 \mu_0 \omega^2 \sum_{j \ne i} \mathbf{G}(\mathbf{r}_i, \mathbf{r}_j) \cdot \mathbf{r}_j(t) +\sum_{j \ne i}\left ((e(\vec{\mathbf r}_j -  \vec{\mathbf  r}_i)\cdot\nabla)\mathbf{E}_{\text{ext}}(\mathbf{r}_{j}, t)+ HO\right)
\end{equation}

This total coupling force will insist the dipole system to be synchronized to have oscillation with common diagonalized modes. Thus, after saturation of the synchronization, the sudden change in the 2D spatial field distribution can't change the synchronized oscillation modes. Hence, the centroid shift will slow down. The final representation of the Lorentz dynamics including gradient forces should be presented as follows.

\begin{eqnarray}
    && m \ddot{\mathbf{r}}_i + m \gamma \dot{\mathbf{r}}_i + m \omega_0^2 \mathbf{r}_i + \beta_i |\mathbf{r}_i| \mathbf{r}_i + \alpha_i |\mathbf{r}_i|^2 \mathbf{r}_i \nonumber\\ &&= -e \mathbf{E}_{\text{ext}}(\mathbf{r}_i, t) + e^2 \mu_0 \omega^2 \sum_{j \ne i} \mathbf{G}(\mathbf{r}_i, \mathbf{r}_j) \cdot \mathbf{r}_j(t) +\sum_{j \ne i}\left ((e(\vec{\mathbf r}_j -  \vec{\mathbf  r}_i)\cdot\nabla)\mathbf{E}_{\text{ext}}(\mathbf{r}_{j}, t)+ HO\right)
\end{eqnarray}

\subsection{Diagonalization of Coupled Dynamical System}
The presence of coupling in the dipole dynamical system contains constraints which modifies the degrees of freedom. Hence, coupling coefficient forms a non-diagonal transformation metric system, and the conventional Cartesian system can't act as an orthonormal frame for such dynamical system. This system can't produce independent dynamical equations and hence, diagonalization is needed to apply on the coupled dipole dynamical equations. This diagonalization establishes the orthonormal frame of reference and modified degrees of freedom. The diagonalization of the dynamics produces coupled-mode oscillatory dipole dynamics which reduces the chaotic nature of the initial random dipole moments. we extend the diagonalization and modal-expansion framework developed earlier section. We demonstrate how the inclusion of gradient and higher-order terms modifies the modal dynamics and consequently the output field. We define the collective displacement vector as follows.

\begin{equation}
R(t)=\big[r_1(t),r_2(t),\dots,r_N(t)\big]^T,
\end{equation}

and the dipole–dipole interaction matrix

\begin{equation}
C_{ij} = 
\begin{cases}
-\,k_0^2\mu_0\omega^2 G(r_i,r_j), & i\neq j, \\
0, & i=j.
\end{cases}
\end{equation}

The effective stiffness matrix becomes

\begin{equation}
K_{\rm eff}=\omega_0^2 I + \frac{e^2}{m}C.
\end{equation}

The nonlinear restoring terms are written as

\begin{equation}
B_2[R]R = [\beta_i|r_i|r_i], \qquad B_3[R]R = [\alpha_i|r_i|^2r_i],
\end{equation}

and the gradient contribution is collected in the source vector

\begin{equation}
F_{\rm grad}(t)=\big[F_{{\rm grad},1}(t),\dots,F_{{\rm grad},N}(t)\big]^T,
\end{equation}

with

\begin{equation}
F_{{\rm grad},i}(t)=\sum_{j\ne i}\left((e(r_j-r_i)\cdot\nabla)E_{\rm ext}(r_j,t)+{\rm HO}\right)_i.
\end{equation}

Thus, the dynamical equation takes the compact vector form

\begin{equation}
\ddot R + \gamma\dot R + K_{\rm eff}R + \frac{1}{m}B_2[R]R + \frac{1}{m}B_3[R]R
= -\frac{e}{m}E_{\rm ext}(t)+\frac{1}{m}F_{\rm grad}(t).
\label{eq:vectorform}
\end{equation}

Assuming harmonic steady state,

\begin{equation}
R(t)=R_\omega e^{-i\omega t}+{\rm c.c.},\qquad 
E_{\rm ext}(t)=E_\omega e^{-i\omega t}+{\rm c.c.},
\end{equation}

the nonlinear terms are approximated (keeping resonant contributions) as

\begin{equation}
|r_i|r_i \approx \sqrt{2}|r_{\omega,i}|(r_{\omega,i}e^{-i\omega t}+{\rm c.c.}), 
\qquad 
|r_i|^2r_i \approx 3|r_{\omega,i}|^2(r_{\omega,i}e^{-i\omega t}+{\rm c.c.}).
\end{equation}

Collecting resonant terms, the amplitude equation becomes

\begin{equation}
\Big[-\omega^2I+i\gamma\omega I+K_{\rm eff}+\tfrac{1}{m}B^{(1)}_2+\tfrac{1}{m}B^{(1)}_3\Big]R_\omega
= -\frac{e}{m}E_\omega+\frac{1}{m}F_{{\rm grad},\omega},
\label{eq:freqdomain}
\end{equation}

where $B^{(1)}_2=\mathrm{diag}(\beta_i\sqrt{2}|r_{\omega,i}|)$ and $B^{(1)}_3=\mathrm{diag}(3\alpha_i|r_{\omega,i}|^2)$. Diagonalize the stiffness matrix:

\begin{equation}
K_{\rm eff}=U\Lambda U^{-1},\qquad \Lambda=\mathrm{diag}(\Omega_1^2,\dots,\Omega_{3N}^2),
\end{equation}

and transform to modal coordinates

\begin{equation}
Q_\omega = U^{-1}R_\omega.
\end{equation}

Equation~\eqref{eq:freqdomain} becomes

\begin{equation}
\Big[-\omega^2I+i\gamma\omega I+\Lambda+\tfrac{1}{m}\widetilde B\Big]Q_\omega
= -\frac{e}{m}U^{-1}E_\omega+\frac{1}{m}U^{-1}F_{{\rm grad},\omega},
\label{eq:modal}
\end{equation}

with $\widetilde B=U^{-1}(B^{(1)}_2+B^{(1)}_3)U$. To leading order we neglect $\widetilde B$ (linear response). The zeroth-order modal amplitude for mode $n$ is,

\begin{equation}
Q^{(0)}_n(\omega)=
\frac{-\tfrac{e}{m}\langle\phi_n|E_\omega\rangle+\tfrac{1}{m}\langle\phi_n|F_{{\rm grad},\omega}\rangle}
{\Omega_n^2-\omega^2-i\gamma\omega},
\label{eq:modalzero}
\end{equation}

where $\phi_n$ is the $n$-th eigenvector (column of $U$) and $\langle\phi_n|\cdot\rangle$ denotes the modal projection. Thus the gradient contribution in the dynamical equation enters as an additional source term in the numerator of the modal response. The total polarization is

\begin{equation}
P(r,\omega)=-Ne\sum_n Q_n(\omega)\phi_n(r).
\label{eq:polarization}
\end{equation}

Using the scalar Green’s function

\begin{equation}
G(r,r')=\frac{e^{ik_0|r-r'|}}{4\pi|r-r'|},
\end{equation}

the radiated output field is expressed as

\begin{equation}
E_{\rm out}(r)=k_0^2\varepsilon_0\int\!\!\int 
G(r,r')\,\chi(r',r'';\omega,|E_\omega|)\,E_\omega(r'')\,d^3r''\,d^3r',
\end{equation}

where the effective susceptibility kernel contains the nonlinear and modal corrections. In modal form, inserting Eq.~\eqref{eq:modalzero}, we obtain

\begin{equation}
E_{\rm out}(r)=
k_0^2\varepsilon_0\sum_n \phi_n(r)\,
\frac{Ne^2}{\varepsilon_0 m}\,
\frac{\langle\phi_n|E_\omega\rangle-\tfrac{1}{e}\langle\phi_n|F_{{\rm grad},\omega}\rangle}
{\Omega_n^2-\omega^2-i\gamma\omega}
\;\otimes\;
\Big(\int\!\!\int \phi_n(r')\otimes\phi_n^*(r'')\,G(r,r')\,E_\omega(r'')\,d^3r''\,d^3r'\Big).
\label{eq:outputfinal}
\end{equation}

Equation~\eqref{eq:outputfinal} shows that the gradient terms in Eq.~\eqref{eq:eq15} act as additional driving sources for the collective dipole modes. They do not primarily shift modal resonances (denominator), but rather enhance or suppress excitation of specific modes via the modal projection $\langle\phi_n|F_{{\rm grad},\omega}\rangle$. This modifies the polarization spectrum and, consequently, the scattered output field. Nonlinear corrections from $B^{(1)}_2$ and $B^{(1)}_3$ Further modify the response in higher-order perturbation theory.

\subsection{Lorentz Force Impact on Dipole Moment}
After synchronisation, if the intensity distribution changes suddenly, the Lorentz Force also comes into play due to the pre-existence of oscillation dipoles. The magnetic force part of the Lorentz force can again perturb the coupling forces. The general mathematical form of Lorentz forces is as follows.

\begin{equation}
    \mathbf F_{Lorentz} = (\mathbf p_i\cdot\nabla)\mathbf E'_{Ext}(\mathbf r_i, t) + \dot{\mathbf p}_i\times \mathbf B (\mathbf r_i,t) + HOMP
\end{equation}

Here, the effect of magnetic field has been considered into account. Although, that effect can be neglected as because $\mathbf{B}$ can be substituted with $\frac{\mathbf{E}}{\mathbf c}$. hence, the numerical impact is very small and hence, can be ignored. $\mathbf{E'}_{Ext}$ is the changed electric field distribution that is redristributed from $\mathbf{E}_{ext}$ due to presence of dynamic turbulence. $HOMP$ is the higher ordered multipole (HOMP) components on Lorentz force. Here, in present context, we have only considered dipoles. Thus, higher orders can be neglected. The force due to new distributed field gives perturbation on the previous synchronized dipole system and thus, tries to re-build the randomness into distribution and oscillation modes. Hence, the perturbation forces can be summed as follows when the field distribution changes after synchronization.

\begin{eqnarray}
    &&\mathbf{\delta F}_{Pert} = \mathbf F_{Tot} - \mathbf F'_{Tot} = \mathbf F_{Ext} + \mathbf F_{Coupling} + \mathbf F_{HO} - \mathbf F'_{Ext} - \mathbf F'_{Coupling} - \mathbf F'_{HO} - \mathbf F_{Lorentz} \nonumber\\&&= -e \mathbf{E}_{\text{ext}}(\mathbf{r}_i, t) + e^2 \mu_0 \omega^2 \sum_{j \ne i} \mathbf{G}(\mathbf{r}_i, \mathbf{r}_j) \cdot \mathbf{r}_j(t) +\sum_{j \ne i}\left ((e(\vec{\mathbf r}_j -  \vec{\mathbf  r}_i)\cdot\nabla)\mathbf{E}_{\text{ext}}(\mathbf{r}_{j}, t)+ HO\right)\nonumber\\&& +e \mathbf{E'}_{\text{ext}}(\mathbf{r}_i, t) - e^2 \mu_0 \omega^2 \sum_{j \ne i} \mathbf{G'}(\mathbf{r}_i, \mathbf{r}_j) \cdot \mathbf{r}_j(t) -\sum_{j \ne i}\left ((e(\vec{\mathbf r}_j -  \vec{\mathbf  r}_i)\cdot\nabla)\mathbf{E'}_{\text{ext}}(\mathbf{r}_{j}, t)+ HO\right) \nonumber\\&&- (\mathbf p_i\cdot\nabla)\mathbf E'_{Ext}(\mathbf r_i, t) - \dot{\mathbf p}_i\times \mathbf B' (\mathbf r_i,t) - HOMP'
\end{eqnarray}

The influence of such gradient force opposes the impact of a sudden external field distribution on synchronised dipole moments. At the same time, such a gradient force reduces the gradient output field through synchronisation of the medium dipole moment. Thus, the output field distribution standard deviations become higher than the input field. If the length of the medium increases, the synchronisation will be more and the standard deviations increases. The fluctuations of the standard deviations due to continuous change in input field distribution will be decreased due to such gradient forces. 

\subsection{d’Alembert’s Principle \& Effective Force of Inertia}
In the present context of electric field force application on dipole system, the presence of dynamic turbulence induces a change in the magnitude and spatial distribution of the optical forces. The change in electric field force induces changes in inertia forces on dipoles. These phenomenon must be discussed using d'Alembert's principle of Lorentz dipole dynamics. For the present dynamics, we can find the following relations from d'Alembert's principle in the presence of two different electric field distributions.

\begin{equation}
    \int{\delta W} = \int{\left ( \mathbf F_{0th} + \mathbf F_{HO} + \mathbf F_{Coupling} + \mathbf F_{Inertia} \right )\cdot dx} = 0 
\end{equation}

\begin{equation}
    \int{\delta W'} = \int{\left ( \mathbf F'_{0th} + \mathbf F'_{HO} + \mathbf F'_{Coupling} + \mathbf F_{Lorentz} + \mathbf F'_{Inertia} \right )\cdot dx} = 0 
\end{equation}

Here, the $(')-$ notation represents the parameters of the changed forced condition. For every field distribution variation and propagation through a medium, the dipole system occupies variable inertia forces. Each inertia force produces a corresponding inertia level in the system where dipoles oscillate. After the propagation of the first field, the dipoles start to oscillate with a coupled mode. After synchronisation, when the field distribution changed, the inertia force from the first field opposes the second field. Thus, the perturbing field comes from the difference between two inertia forces for two electric fields with different distributions. Hence, the perturbation force due to the presence of dynamic turbulence influenced the variable electric fields can be discussed as follows.

\begin{equation}
    \delta \mathbf{F}_{Pert} =  \mathbf F'_{Inertia} - \mathbf F_{Inertia}
\end{equation}

In the present context, the presence of dynamic turbulence impact on the electric field spatial distribution of the propagating optical field includes the time dependency of this perturbation force. Hence, we must have $\delta \mathbf F_{Pert}\rightarrow\delta\mathbf F_{Pert}(t)$. After such perturbation, the dynamical equation must look like the following equation.

\begin{eqnarray}
    && m \ddot{\mathbf{r}}_i + m \gamma \dot{\mathbf{r}}_i + m \omega_0^2 \mathbf{r}_i + \beta_i |\mathbf{r}_i| \mathbf{r}_i + \alpha_i |\mathbf{r}_i|^2 \mathbf{r}_i \nonumber\\ &&= -e \mathbf{E}_{\text{ext}}(\mathbf{r}_i, t) + e^2 \mu_0 \omega^2 \sum_{j \ne i} \mathbf{G}(\mathbf{r}_i, \mathbf{r}_j) \cdot \mathbf{r}_j(t) +\sum_{j \ne i}\left ((e(\vec{\mathbf r}_j -  \vec{\mathbf  r}_i)\cdot\nabla)\mathbf{E}_{\text{ext}}(\mathbf{r}_{j}, t)+ HO\right) +  \delta \mathbf{F}_{Pert} (t)
\end{eqnarray}

We can find the following conditions on the perturbed force depending upon corresponding magnitudes.
\begin{itemize}
    \item $\delta\mathbf F_{Pert}\rightarrow 0$: The output field distribution can be caused by the saturated synchronized dipole moment distribution. Hence, the turbulence impact can be compensated fully with presence of medium dipole-dipole coupling energy transitions.
    \item $\delta\mathbf F_{Pert}\rightarrow $Small but $\neq 0$: The dynamic nature of turbulence changes the synchronization with induction of perturbed inertia force. The perturbation is small, hence, output field turbulence impact can be found compensated.
    \item $\delta\mathbf F_{Pert}>0$: In such case, the output field distribution will be dependent upon the frequency of change of perturbation. If turbulence is strong, i.e., change of perturbation is rapid, the medium dipole coupled system can't find time to be synchronized. Thus, for strong turbulence, we can find un-compensated turbulence impacted output field. If the turbulence is weak, output field can be found compensated from turbulence.
\end{itemize}

\section{Statistical Background}\label{3}
In this section we provide a detailed theoretical framework for analyzing higher-order 
statistics of two-dimensional intensity images using cumulants. The method begins 
with the interpretation of intensity values as a probability measure, proceeds through 
mean and covariance estimation, performs Cholesky whitening for standardization, 
and finally computes higher-order cumulants and employs a Gram--Charlier expansion 
to model deviations from Gaussianity. 

\subsection{Probability Measure from Image Intensity}

Consider a two-dimensional intensity distribution $I_{ij}\geq 0$ defined on discrete 
pixels with coordinates $(x_j,y_i)$. We interpret intensities as defining a probability 
measure after normalization. Define weighted intensities $W_{ij}$ (which may include 
masking) and the normalization constant,

\begin{equation}
S=\sum_{i,j}W_{ij}, \qquad p_{ij}=\frac{W_{ij}}{S},
\label{eq:probmeasure}
\end{equation}

such that $\sum_{i,j}p_{ij}=1$. For any function $f(X,Y)$ of spatial coordinates, the intensity-weighted expectation is,

\begin{equation}
\mathbb{E}[f(X,Y)]=\sum_{i,j}p_{ij} f(x_j,y_i).
\end{equation}

This constructs a discrete probability distribution proportional to intensity.

\subsection{Mean and Covariance}

The centroid of the distribution is given by

\begin{equation}
\mu_x=\mathbb{E}[X], \qquad \mu_y=\mathbb{E}[Y].
\end{equation}

The covariance matrix of second-order statistics is,

\begin{equation}
\Sigma=
\begin{pmatrix}
\sigma_{xx} & \sigma_{xy}\\
\sigma_{xy} & \sigma_{yy}
\end{pmatrix}
=
\begin{pmatrix}
\mathbb{E}[(X-\mu_x)^2] & \mathbb{E}[(X-\mu_x)(Y-\mu_y)] \\
\mathbb{E}[(X-\mu_x)(Y-\mu_y)] & \mathbb{E}[(Y-\mu_y)^2]
\end{pmatrix}.
\label{eq:covmatrix}
\end{equation}

\subsection{Cholesky Whitening}

To remove correlations and anisotropic scaling, we employ Cholesky whitening.  
Since $\Sigma$ is symmetric positive definite, it admits a unique Cholesky factorization:

\begin{equation}
\Sigma=LL^\top,
\end{equation}

where $L$ is lower triangular. Standardized (whitened) coordinates are defined by

\begin{equation}
z=\begin{pmatrix}z_1\\z_2\end{pmatrix}=L^{-1}\begin{pmatrix}x-\mu_x \\ y-\mu_y\end{pmatrix}.
\end{equation}

By construction,

\begin{equation}
\mathbb{E}[z]=0, \qquad \mathrm{Cov}(z)=I_2,
\end{equation}

so that Gaussian structure is fully normalized, and deviations from Gaussianity appear only in higher cumulants. Cholesky whitening has advantages over eigen-decomposition whitening: it avoids eigenvector sorting and sign ambiguities, is unique, and is numerically stable.

\subsection{Standardized Moments}

Using whitened samples $(z_{1k},z_{2k})$ with weights $w_k=p_{ij}$, the standardized 
moments are defined as,

\begin{equation}
m_{pq}=\sum_k w_k z_{1k}^p z_{2k}^q.
\end{equation}

From whitening constraints:

\begin{equation}
m_{20}=m_{02}=1,\qquad m_{11}=0.
\end{equation}

Higher-order standardized moments include,

\begin{align}
& m_{30}, m_{21}, m_{12}, m_{03}, \\
& m_{40}, m_{31}, m_{22}, m_{13}, m_{04}.
\end{align}

\subsection{Cumulants from Moments}

The cumulant generating function is,

\begin{equation}
K(t)=\log \mathbb{E}[e^{t_1z_1+t_2z_2}],
\end{equation}

whose Taylor coefficients give the cumulants. For standardized variables we have, 

\paragraph{Third-order cumulants (skewness):}
\begin{equation}
k_{30}=m_{30},\qquad k_{21}=m_{21},\qquad 
k_{12}=m_{12},\qquad k_{03}=m_{03}.
\end{equation}

\paragraph{Fourth-order cumulants (kurtosis excess):}
\begin{align}
k_{40} &= m_{40}-3m_{20}^2, \\
k_{04} &= m_{04}-3m_{02}^2, \\
k_{31} &= m_{31}-3m_{20}m_{11}, \\
k_{13} &= m_{13}-3m_{02}m_{11}, \\
k_{22} &= m_{22}-m_{20}m_{02}-2m_{11}^2.
\end{align}

Using $m_{20}=m_{02}=1$ and $m_{11}=0$, these simplify to,

\begin{equation}
k_{40}=m_{40}-3,\quad k_{04}=m_{04}-3,\quad 
k_{31}=m_{31},\quad k_{13}=m_{13},\quad k_{22}=m_{22}-1.
\end{equation}

\subsection{Norms of Skewness and Kurtosis}

To quantify overall non-Gaussianity, Euclidean norms of cumulant vectors are used:

\begin{align}
|\text{skew}|_3 &= \sqrt{k_{30}^2+k_{21}^2+k_{12}^2+k_{03}^2}, \\
|\text{kurt}|_4 &= \sqrt{k_{40}^2+k_{31}^2+k_{22}^2+k_{13}^2+k_{04}^2}.
\end{align}

\subsection{Gram--Charlier Expansion}

Let $\phi(r)$ be the Gaussian density with mean $\mu$ and covariance $\Sigma$:

\begin{equation}
\phi(r)=\frac{1}{2\pi\sqrt{\det \Sigma}}
\exp\!\Big(-\tfrac{1}{2}(r-\mu)^\top \Sigma^{-1}(r-\mu)\Big).
\end{equation}

In whitened coordinates $z=L^{-1}(r-\mu)$,

\begin{equation}
\phi(z)=(2\pi)^{-1}\exp\!\left(-\tfrac{1}{2}|z|^2\right).
\end{equation}

The multivariate Hermite polynomials are defined by,

\begin{equation}
H_\alpha(z)=(-1)^{|\alpha|}\phi(z)^{-1}\partial^\alpha \phi(z),\qquad \alpha=(\alpha_1,\alpha_2).
\end{equation}

Up to fourth order:

\begin{align}
&H_{30}=z_1^3-3z_1, \quad H_{21}=z_1^2z_2-z_2, \quad 
H_{12}=z_1z_2^2-z_1, \quad H_{03}=z_2^3-3z_2, \\
&H_{40}=z_1^4-6z_1^2+3, \quad H_{31}=z_1^3z_2-3z_1z_2, \quad 
H_{22}=z_1^2z_2^2-z_1^2-z_2^2+1, \\
&H_{13}=z_1z_2^3-3z_1z_2,\quad H_{04}=z_2^4-6z_2^2+3.
\end{align}

The Gram--Charlier expansion of the density is,

\begin{align}
p(r) &\approx \phi(r)\Bigg[1+\frac{1}{6}\big(k_{30}H_{30}+3k_{21}H_{21}+3k_{12}H_{12}+k_{03}H_{03}\big) \nonumber\\
&\quad+\frac{1}{24}\big(k_{40}H_{40}+4k_{31}H_{31}+6k_{22}H_{22}+4k_{13}H_{13}+k_{04}H_{04}\big)\Bigg].
\end{align}

\subsection{Fitting to Observed Intensity}

To match the expansion to measured image intensity, we include a scaling and offset:

\begin{equation}
I_{\rm fit}(r)=a\,p(r)+b,
\end{equation}

with $a,b$ determined via least-squares fitting to account for global amplitude and background. This procedure yields a rigorous method to extract higher-order statistical features (skewness and kurtosis excess) from 2D intensity images. Whitening ensures that second-order effects are normalized, isolating true higher-order non-Gaussian characteristics of the data. The complete analysis pipeline proceeds as follows. First, the image intensity is normalized in order to define a probability distribution. From this distribution, the mean $\mu$ and covariance matrix $\Sigma$ are computed. Cholesky whitening is then applied to remove correlations and obtain standardized coordinates $z$. Using these whitened coordinates, standardized moments are evaluated and subsequently converted into cumulants. The skewness and kurtosis norms are then quantified to measure the degree of non-Gaussianity. These cumulants are incorporated into a Gram--Charlier expansion, which provides a model for the intensity distribution beyond the Gaussian approximation. Finally, the expansion is fitted to the observed image intensity by including an overall scale and offset, thereby accounting for amplitude variations and background contributions.

\subsection{Fitted Power variation due to Dynamic Turbulence}
In the present theoretical frame work we have fitted our 2D optical beam images using 2D bi-variate Gaussian function with skewness and Kurtosis as noise distortions. After dynamic turbulence impact, the gaussian beam spatial distribution get expanded. The centroid starts to move. Hence, for every frame of measurements, the images become unbound in structures. Thus, fitted function bounded volume can become an important parameter to identify the structural changes in each frame in presence of dynamic turbulence impact. The volume can be formulized as follows.

\begin{equation}
    V_{Frame}=\int_{\mathbb R^2} I_{\rm fit}(\mathbf r,t)\,d^2\mathbf r=2\pi I_{Max}\sqrt{|\Sigma|}
\end{equation}

The volume found from the above equation can't give total power of each image frame. It is the power for corresponding fitted 2D functions under pre-defined orders of distortion i.e., the Skewness and Kurtosis. As the higher orders are limited to Skewness and Kurtosis, the volume under those fitted functions will not be same as the total power for each frame. The change of the volume with change in frame must be defined by the impact of dynamic turbulence. Hence, the Volumes of those fitted functions must be an unified parameter to identify the turbulence.

\subsection{Volume Scintillation as Generalization of Pointwise Scintillation}
In this section, we present a rigorous derivation of the relationship between the integrated-volume scintillation index \(\sigma_V^2\) and the pointwise (local) scintillation index \(\sigma_I^2(\mathbf{r})\). We assumed minimal assumptions about the optical field, derived an exact identity linking the two indices, reformulated it using the local correlation coefficient, specialized the results to Gaussian random optical fields via the cross-spectral density, and concluded with useful inequalities derived from the Cauchy-Schwarz principle. This link is fundamental in understanding how spatial integration over a detector area reduces scintillation effects compared to pointwise measurements, particularly in contexts like atmospheric turbulence or light-matter interactions. A transverse plane has been assumed \(\mathbb{R}^2\) where the optical field \(E(\mathbf{r}, t)\) is defined such that all second- and fourth-order moments of the intensity exist and are finite. The intensity is given by \(I(\mathbf{r}, t) = \frac{|E(\mathbf{r}, t)|^2}{\max(|E(\mathbf{r}, t)|^2)}\), ensuring it is measurable and integrable. Ensemble averages are denoted by \(\langle \cdot \rangle\) (or time averages under ergodicity). Key notations include the instantaneous integrated volume \(V_{\text{inst}}(t) = \int_{\mathbb{R}^2} I(\mathbf{r}, t) \, d^2\mathbf{r}\), the mean integrated volume \(\langle V \rangle = \int_{\mathbb{R}^2} \langle I(\mathbf{r}) \rangle \, d^2\mathbf{r}\), the covariance \(\text{Cov}(I_1, I_2) = \langle I(\mathbf{r}_1) I(\mathbf{r}_2) \rangle - \langle I(\mathbf{r}_1) \rangle \langle I(\mathbf{r}_2) \rangle\), and the pointwise scintillation \(\sigma_I^2(\mathbf{r}) = \frac{\text{Var}[I(\mathbf{r})]}{\langle I(\mathbf{r}) \rangle^2}\). The derivation starts with the definition of the integrated-volume scintillation index:

\begin{equation}
    \sigma_V^2 = \frac{\text{Var}[V_{\text{inst}}]}{\langle V \rangle^2}
\end{equation}

Expanding the variance yields:

\begin{equation}
    \text{Var}[V_{\text{inst}}] = \left\langle \left( \int_{\mathbb{R}^2} I(\mathbf{r}_1) \, d^2\mathbf{r}_1 \right)^2 \right\rangle - \langle V \rangle^2 = \iint_{\mathbb{R}^2} \langle I(\mathbf{r}_1) I(\mathbf{r}_2) \rangle \, d^2\mathbf{r}_1 d^2\mathbf{r}_2 - \langle V \rangle^2.
\end{equation}

This simplifies to:

\begin{equation}
    \text{Var}[V_{\text{inst}}] = \iint_{\mathbb{R}^2} \text{Cov}(I(\mathbf{r}_1), I(\mathbf{r}_2)) \, d^2\mathbf{r}_1 d^2\mathbf{r}_2
\end{equation}

since \(\iint_{\mathbb{R}^2} \langle I(\mathbf{r}_1) \rangle \langle I(\mathbf{r}_2) \rangle \, d^2\mathbf{r}_1 d^2\mathbf{r}_2 = \langle V \rangle^2\). Thus, the exact identity is:

\begin{equation}
    \sigma_V^2 = \frac{1}{\langle V \rangle^2} \iint_{\mathbb{R}^2} \text{Cov}(I(\mathbf{r}_1), I(\mathbf{r}_2)) \, d^2\mathbf{r}_1 d^2\mathbf{r}_2
\end{equation}

This relation shows that \(\sigma_V^2\) is the double spatial integral of the intensity covariance, normalized by the square of the mean integrated power. It is exact and holds without additional assumptions beyond the minimal ones stated. To connect explicitly to the pointwise scintillation, the normalized covariance is introduced:

\begin{equation}
    C(\mathbf{r}_1, \mathbf{r}_2) = \frac{\text{Cov}(I_1, I_2)}{\langle I_1 \rangle \langle I_2 \rangle}
\end{equation}

so \(\text{Cov}(I_1, I_2) = \langle I_1 \rangle \langle I_2 \rangle C(\mathbf{r}_1, \mathbf{r}_2)\). Substituting gives:

\begin{equation}
    \sigma_V^2 = \frac{1}{\langle V \rangle^2} \iint_{\mathbb{R}^2} \langle I_1 \rangle \langle I_2 \rangle C(\mathbf{r}_1, \mathbf{r}_2) \, d^2\mathbf{r}_1 d^2\mathbf{r}_2
\end{equation}

Alternatively, using the Pearson correlation coefficient:

\begin{equation}
    \rho(\mathbf{r}_1, \mathbf{r}_2) = \frac{\text{Cov}(I_1, I_2)}{\sqrt{\text{Var}[I_1] \text{Var}[I_2]}}
\end{equation}

and noting \(\text{Var}[I(\mathbf{r})] = \sigma_I^2(\mathbf{r}) \langle I(\mathbf{r}) \rangle^2\), we have:

\begin{equation}
    \text{Cov}(I_1, I_2) = \rho(\mathbf{r}_1, \mathbf{r}_2) \langle I_1 \rangle \langle I_2 \rangle \sqrt{\sigma_I^2(\mathbf{r}_1) \sigma_I^2(\mathbf{r}_2)}
\end{equation}

Thus:

\begin{equation}
    \sigma_V^2 = \frac{1}{\langle V \rangle^2} \iint_{\mathbb{R}^2} \rho(\mathbf{r}_1, \mathbf{r}_2) \langle I_1 \rangle \langle I_2 \rangle \sqrt{\sigma_I^2(\mathbf{r}_1) \sigma_I^2(\mathbf{r}_2)} \, d^2\mathbf{r}_1 d^2\mathbf{r}_2
\end{equation}

This expression directly links \(\sigma_V^2\) to the local scintillation \(\sigma_I^2(\mathbf{r})\), the mean intensity profile, and the spatial correlation \(\rho\).

\subsection{Gaussian-Field Specialization}

For zero-mean circular-complex Gaussian random fields, the fourth-order intensity moment factorizes:

\begin{equation}
    \text{Cov}(I_1, I_2) = |\Gamma(\mathbf{r}_1, \mathbf{r}_2, 0)|^2
\end{equation}

where \(\Gamma(\mathbf{r}_1, \mathbf{r}_2, 0) = \langle E^*(\mathbf{r}_1, t) E(\mathbf{r}_2, t) \rangle\) is the mutual coherence at zero time lag. Substituting yields:

\begin{equation}
    \sigma_V^2 = \frac{1}{\langle V \rangle^2} \iint_{\mathbb{R}^2} |\Gamma(\mathbf{r}_1, \mathbf{r}_2, 0)|^2 \, d^2\mathbf{r}_1 d^2\mathbf{r}_2
\end{equation}

Using the cross-spectral density \(W(\mathbf{r}_1, \mathbf{r}_2, \omega)\):

\begin{equation}
    \Gamma(\mathbf{r}_1, \mathbf{r}_2, 0) = \frac{1}{2\pi} \int_{-\infty}^{\infty} W(\mathbf{r}_1, \mathbf{r}_2, \omega) \, d\omega
\end{equation}

the expression becomes:

\begin{equation}
    \sigma_V^2 = \frac{1}{(2\pi)^2 \langle V \rangle^2} \int d\omega \int d\omega' \iint_{\mathbb{R}^2} W(\mathbf{r}_1, \mathbf{r}_2, \omega) W^*(\mathbf{r}_1, \mathbf{r}_2, \omega') \, d^2\mathbf{r}_1 d^2\mathbf{r}_2
\end{equation}

If cross-frequency correlations vanish, it simplifies to:

\begin{equation}
    \sigma_V^2 \approx \frac{1}{2\pi \langle V \rangle^2} \int d\omega \iint_{\mathbb{R}^2} |W(\mathbf{r}_1, \mathbf{r}_2, \omega)|^2 \, d^2\mathbf{r}_1 d^2\mathbf{r}_2
\end{equation}

This specialization provides a closed-form expression in terms of coherence functions, useful for analytical models like Gaussian beams with Gaussian coherence.

\subsection{Inequalities and Bounds}

Two inequalities are derived. First, from Cauchy-Schwarz:

\begin{equation}
    |\text{Cov}(I_1, I_2)| \leq \sqrt{\text{Var}[I_1] \text{Var}[I_2]} = \langle I_1 \rangle \langle I_2 \rangle \sqrt{\sigma_I^2(\mathbf{r}_1) \sigma_I^2(\mathbf{r}_2)}
\end{equation}

leading to:

\begin{equation}
    \sigma_V^2 \leq \left( \frac{\int \langle I(\mathbf{r}) \rangle \sqrt{\sigma_I^2(\mathbf{r})} \, d^2\mathbf{r}}{\langle V \rangle} \right)^2 \leq \max_{\mathbf{r}} \sigma_I^2(\mathbf{r})
\end{equation}

Second, since \(|\rho(\mathbf{r}_1, \mathbf{r}_2)| \leq 1\):

\begin{equation}
    \sigma_V^2 \leq \frac{1}{\langle V \rangle^2} \iint_{\mathbb{R}^2} \langle I_1 \rangle \langle I_2 \rangle \sqrt{\sigma_I^2(\mathbf{r}_1) \sigma_I^2(\mathbf{r}_2)} \, d^2\mathbf{r}_1 d^2\mathbf{r}_2
\end{equation}

These bounds illustrate that \(\sigma_V^2\) is reduced below the pointwise scintillation when the detector area exceeds the coherence area, due to spatial averaging. The exact identity requires no Gaussian assumption and represents an area-weighted double integral of intensity covariance. The explicit link incorporates local scintillation, mean intensity, and correlation. For Gaussian fields, the covariance equals the squared mutual coherence, yielding forms in terms of \(\Gamma\) or \(W\). Analytical evaluations with specific models can further show dependence on detector area and coherence length. The final relations summarize:

\begin{equation}
    \sigma_V^2 = \frac{1}{\langle V \rangle^2} \iint \text{Cov}(I(\mathbf{r}_1), I(\mathbf{r}_2)) \, d^2\mathbf{r}_1 d^2\mathbf{r}_2
\end{equation}
\begin{equation}
    \sigma_V^2 = \frac{1}{\langle V \rangle^2} \iint \rho(\mathbf{r}_1, \mathbf{r}_2) \langle I_1 \rangle \langle I_2 \rangle \sqrt{\sigma_I^2(\mathbf{r}_1) \sigma_I^2(\mathbf{r}_2)} \, d^2\mathbf{r}_1 d^2\mathbf{r}_2
\end{equation}
\begin{equation}
    \sigma_V^2 = \frac{1}{\langle V \rangle^2} \iint |\Gamma(\mathbf{r}_1, \mathbf{r}_2, 0)|^2 \, d^2\mathbf{r}_1 d^2\mathbf{r}_2 = \frac{1}{(2\pi)^2 \langle V \rangle^2} \iint d\omega d\omega' \iint W(\mathbf{r}_1, \mathbf{r}_2, \omega) W^*(\mathbf{r}_1, \mathbf{r}_2, \omega') \, d^2\mathbf{r}_1 d^2\mathbf{r}_2
\end{equation}

\section{Experimental Varifications}\label{4}
\begin{figure}[H]
\centering
\begin{minipage}[b]{0.75\textwidth}
    \includegraphics[width=\textwidth]{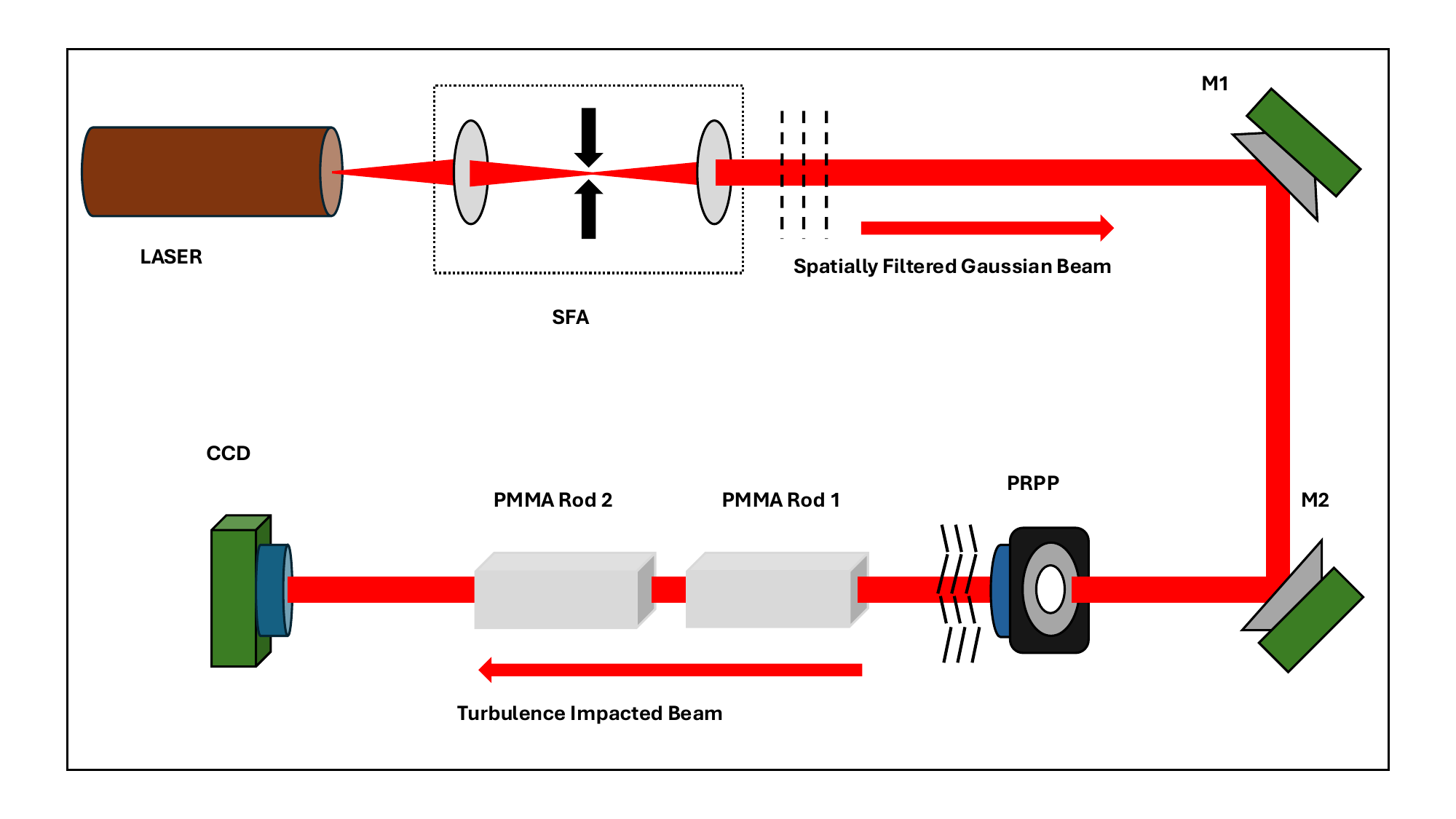}
    \caption{Experimental procedure with 2 PMMA Rod}
    \label{P0}
\end{minipage}
\end{figure}

The experimental arrangement is illustrated schematically in Fig.~\ref{P0}. A continuous-wave laser beam was first passed through a spatial filter assembly (SFA) to generate a clean Gaussian profile with reduced higher-order distortions. The spatially filtered Gaussian beam was directed by two mirrors (M1 and M2) to control propagation and alignment. A programmable rotating phase plate (PRPP) was placed in the optical path to introduce controlled turbulence effects, achieved by imparting dynamic random phase modulations. Following this stage, the turbulence-impacted beam was transmitted through polymethyl methacrylate (PMMA) rods acting as dielectric media. One or two rods were inserted to investigate cumulative light–matter interaction effects under turbulence conditions. The transmitted beam was finally recorded by a charge-coupled device (CCD) camera, which provided turbulence impacted and turbulence impact compensated intensity distributions of the emerging field.

\begin{figure}[h]
\centering
\begin{minipage}[b]{0.75\textwidth}
    \includegraphics[width=\textwidth]{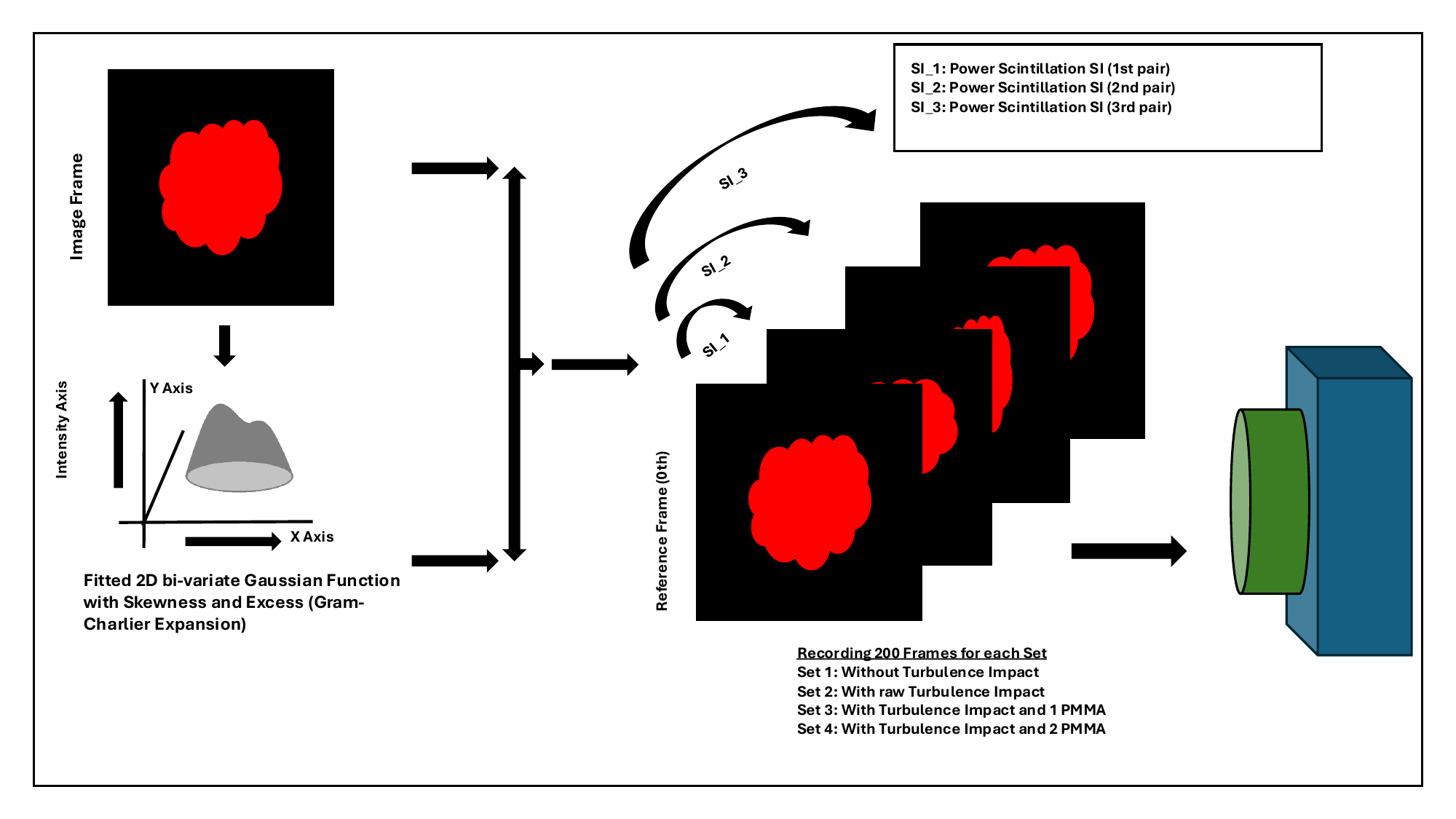}
    \caption{Block diagram Data Analysis Scheme}
    \label{P1}
\end{minipage}
\end{figure}

The data collection scheme, summarized in Fig.~\ref{P1}, was designed to capture statistical fluctuations of the transmitted beam under four different experimental conditions. For each condition, 200 frames were recorded to ensure statistical convergence. The four sets are: (i) Set~1: reference without turbulence (baseline), (ii) Set~2: turbulence introduced via the PRPP without PMMA rods, (iii) Set~3: turbulence combined with one PMMA rod, and (iv) Set~4: turbulence combined with two PMMA rods. Each recorded frame was analyzed by fitting a two-dimensional bivariate Gaussian function to the intensity profile, with higher-order deviations quantified through skewness and kurtosis excess using a Gram-Charlier expansion. This systematic recording and analysis procedure provided direct comparison between turbulence-only and turbulence–PMMA coupled conditions, enabling a robust statistical characterization of light propagation dynamics through dielectric media under turbulence.

Classical turbulence arises from irregular velocity fluctuations in viscous fluids such as the atmosphere, where flow can exist in either laminar or turbulent states. Laminar flow is smooth and orderly, while turbulence is dominated by random subflows, or eddies, that enhance mixing. The transition between these regimes is governed by the Reynolds number, $Re = Vl/\nu$, where $V$ is velocity, $l$ the flow scale, and $\nu$ the kinematic viscosity. When $Re$ exceeds a critical value (typically $\sim 10^5$ near the ground), turbulence develops. Kolmogorov’s theory describes turbulence as statistically homogeneous and isotropic at small scales, with large-scale energy generated by shear or convection cascading down to smaller eddies. Energy transfer occurs across an inertial range bounded by an outer scale $L_0$ and an inner scale $l_0$, until dissipation converts residual energy into heat. The corresponding turbulence power spectrum is expressed as
\begin{equation}
\Phi(k)=0.023r_0^{-5/3}k^{-11/3}.
\end{equation}
The Pseudo Random Phase Plate (PRPP) used in our experiment is a five-layered optical device designed to replicate atmospheric turbulence. It consists of two BK7 glass windows enclosing a central acrylic layer imprinted with a Kolmogorov-type turbulence profile. Near-index-matching polymer layers on each side of the acrylic provide mechanical stability, while the glass sealing enhances durability and minimizes environmental effects. The plate, about 10 mm thick, is robust and easily mounted on a rotary stage. It generates aberrated wavefronts with adjustable Fried coherence lengths ($r_0$ = 16–32 samples) across 4096 phase points, enabling controlled turbulence simulations.

\section{Results Analysis and Discussion}\label{5}
\begin{figure}[H]
\centering
\begin{minipage}[b]{0.75\textwidth}
    \includegraphics[width=\textwidth]{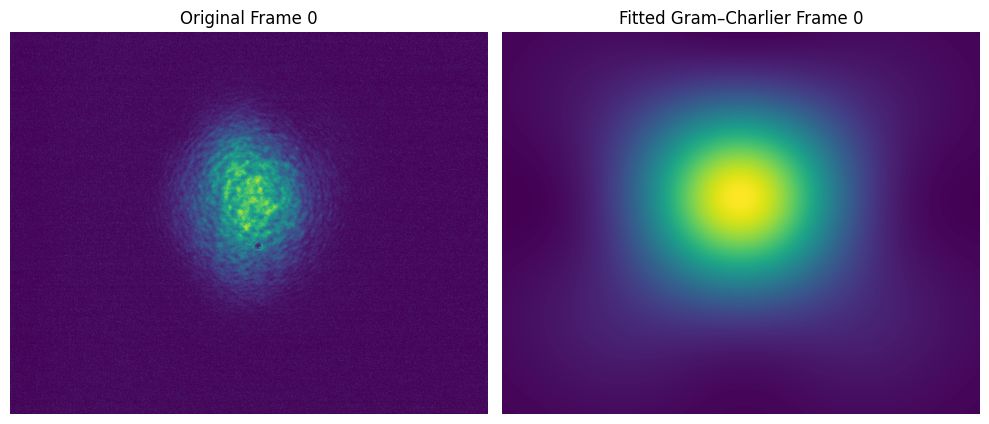}
    \caption{Original Image and Corresponding Fitted Image for Turbulence-free Beam}
    \label{R1a}
\end{minipage}
\end{figure}

\begin{figure}[H]
\centering
\begin{minipage}[b]{0.75\textwidth}
    \includegraphics[width=\textwidth]{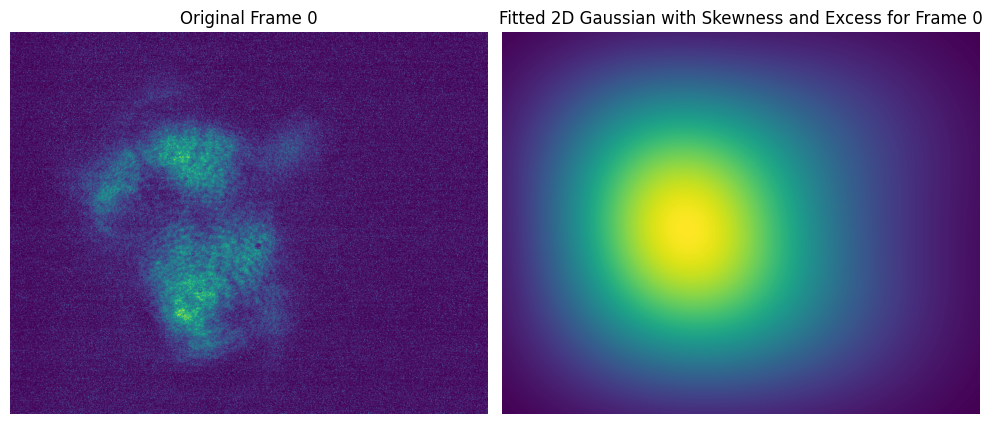}
    \caption{Original Image and Corresponding Fitted Image for Set 1: Raw Turbulence}
    \label{R1b}
\end{minipage}
\end{figure}

\begin{figure}[H]
\centering
\begin{minipage}[b]{0.75\textwidth}
    \includegraphics[width=\textwidth]{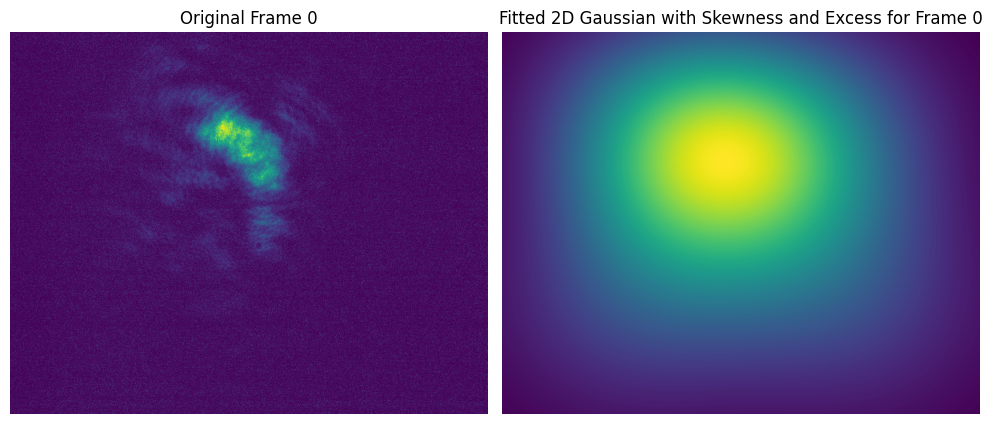}
    \caption{Original Image and Corresponding Fitted Image for Set 2: Turbulence with 1 PMMA Rod}
    \label{R1c}
\end{minipage}
\end{figure}

\begin{figure}[H]
\centering
\begin{minipage}[b]{0.75\textwidth}
    \includegraphics[width=\textwidth]{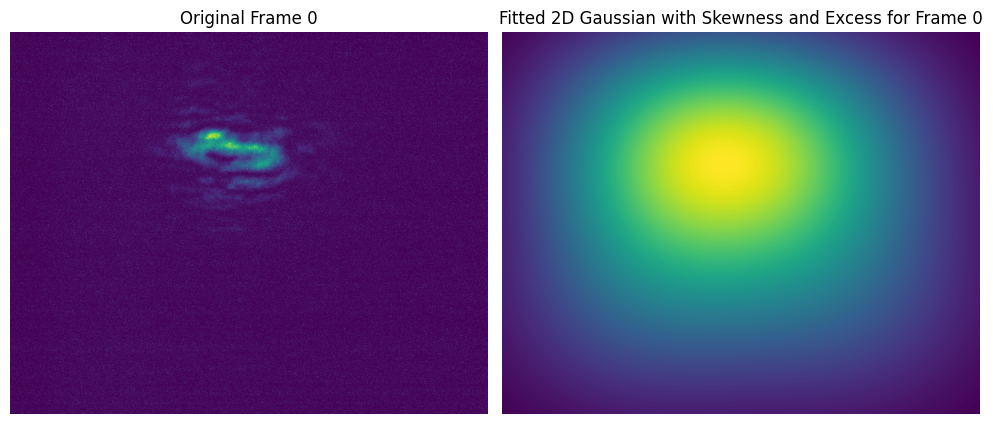}
    \caption{Original Image and Corresponding Fitted Image for Set 3: Turbulence with 2 PMMA Rod}
    \label{R1d}
\end{minipage}
\end{figure}

The recorded images were analyzed by fitting a mathematical model to the experimental intensity distributions of the laser beam. In each figure, the left panel displays the raw experimental image, representing the observed spatial cross-section of the laser beam under the given propagation condition. The right panel displays the corresponding fitted image, which is generated from a model function that approximates the experimental intensity distribution. The fitting was performed using either a two-dimensional Gaussian profile or its extended form via a Gram-Charlier expansion, which incorporates higher-order corrections in the form of skewness (third-order cumulants) and kurtosis excess (fourth-order cumulants). This approach allows the fitted image to capture not only the central Gaussian-like behavior of the beam but also the deviations arising from turbulence-induced distortions. The comparative results across the three experimental sets highlight the role of PMMA rods in compensating turbulence effects. The first figure, labeled (Figure: \ref{R1a}), shows a relatively smooth, symmetric, and turbulence-free beam. The fitted Gram-Charlier function accurately captures this shape, resulting in a clean, circular profile. The beam's intensity distribution closely resembles a standard Gaussian. In Set~S1 (Figure: \ref{R1b}), only turbulence was introduced by the programmable rotating phase plate (PRPP), and no PMMA rods were placed in the beam path. The raw images in this case show significant distortion and spread, which is reflected in the fitted models as enhanced skewness and non-zero kurtosis excess. In Set~S2 (Figure: \ref{R1c}), a single PMMA rod was inserted into the beam path. The fitted results for this configuration indicate a partial compensation of turbulence-induced asymmetries: the skewness values are reduced, and the fitted profiles more closely resemble the ideal Gaussian shape, although some residual excess kurtosis remains. Finally, in Set~S3 (Figure: \ref{R1d}), two PMMA rods were placed sequentially in the optical path. The resulting fitted images show a further suppression of turbulence effects, with the intensity distributions approaching near-Gaussian behavior. In this case, both skewness and kurtosis excess are significantly reduced, indicating that the presence of two rods provides enhanced compensation against turbulence distortions. Overall, the side-by-side comparison of experimental images and fitted models demonstrates the effectiveness of the Gram-Charlier based fitting procedure in quantifying higher-order deviations, as well as the physical compensating role of PMMA rods in mitigating turbulence-induced perturbations in the spatial profile of the transmitted laser beam.

% Results and Fittings


\subsection{Centroid Shift Compensation \& Presence of Inertia Force}
\begin{figure}[H]
        \centering
        \begin{subfigure}[t]{0.48\textwidth}
            \centering
            \includegraphics[width=\linewidth]{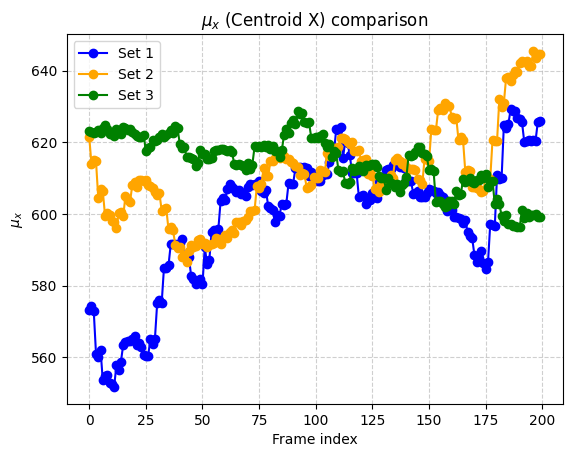}
            \caption{Centroid along 'X' Axis}
        \end{subfigure}
        \hfill
        \begin{subfigure}[t]{0.48\textwidth}
            \centering
            \includegraphics[width=\linewidth]{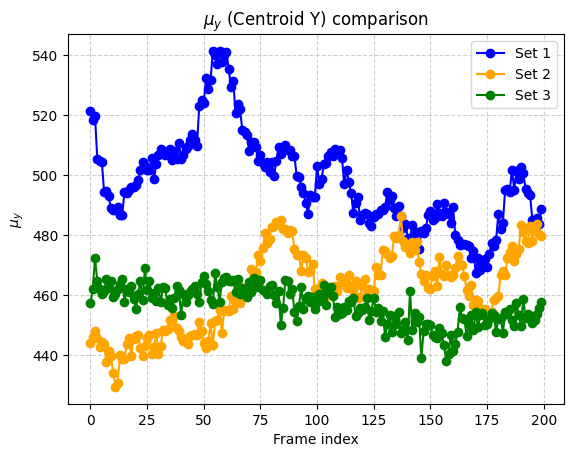}
            \caption{Centroid along 'Y' Axis}
        \end{subfigure}
        \caption{Centroids for three cases. Set 1: Raw Turbulence, Set 2: Turbulence with 1 PMMA Rod, Set 3: Turbulence with 2 PMMA Rod.}
        \label{Cent1}
    \end{figure}

\begin{figure}[H]
        \centering
        \begin{subfigure}[t]{0.48\textwidth}
            \centering
            \includegraphics[width=\linewidth]{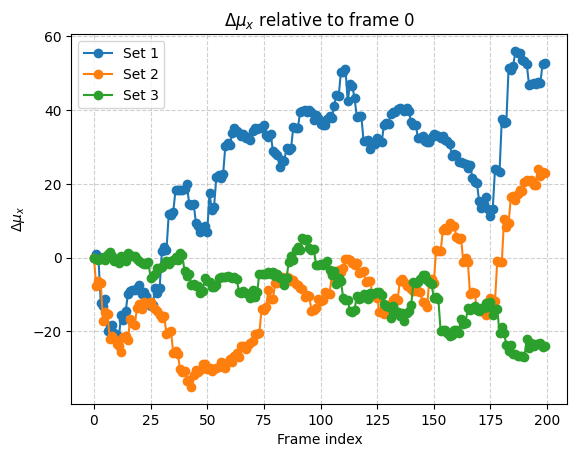}
            \caption{Relative Centroid Shift along 'X' Axis}
        \end{subfigure}
        \hfill
        \begin{subfigure}[t]{0.48\textwidth}
            \centering
            \includegraphics[width=\linewidth]{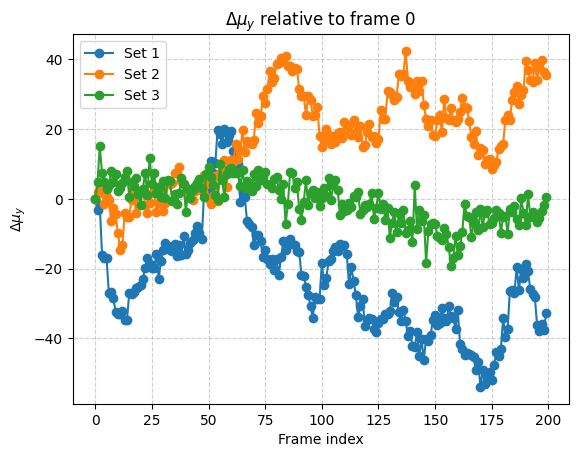}
            \caption{Relative Centroid Shift along 'Y' Axis}
        \end{subfigure}
        \caption{Relative Centroids shifts for three cases. Set 1: Raw Turbulence, Set 2: Turbulence with 1 PMMA Rod, Set 3: Turbulence with 2 PMMA Rod.}
        \label{Cent2}
    \end{figure}

A quantitative analysis of the beam centroid was carried out to assess stability across three different experimental conditions: (i) raw turbulence without PMMA rods (Set~1), (ii) turbulence with a single PMMA rod (Set~2), and (iii) turbulence with two PMMA rods (Set~3). The centroid positions along the $x$- and $y$-axes, denoted by $(\mu_x, \mu_y)$, were tracked over 200 recorded frames (Figure: \ref{Cent1}). In addition, the relative shifts with respect to the reference frame, defined as $(\Delta \mu_x, \Delta \mu_y)=(\mu_x-\mu_x^{(0)},\mu_y-\mu_y^{(0)})$, were computed to highlight the temporal deviations caused by turbulence and medium interaction (Figure: \ref{Cent2}).

% Centroids

\begin{figure}[H]
        \centering
        \begin{subfigure}[t]{0.30\textwidth}
            \centering
            \includegraphics[width=\linewidth]{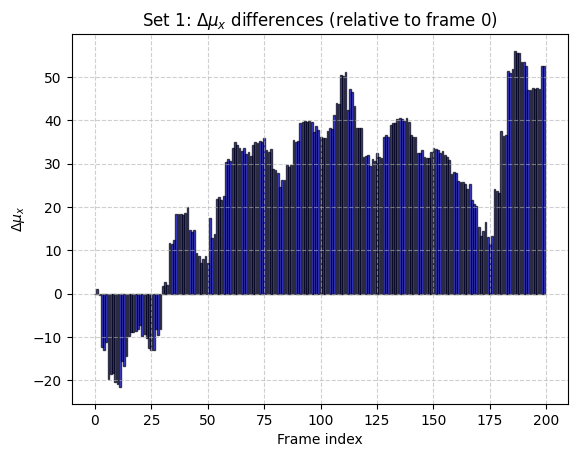}
            \caption{Relative Centroid Shift along 'X' Axis for Set 1}
        \end{subfigure}
        \hfill
        \begin{subfigure}[t]{0.30\textwidth}
            \centering
            \includegraphics[width=\linewidth]{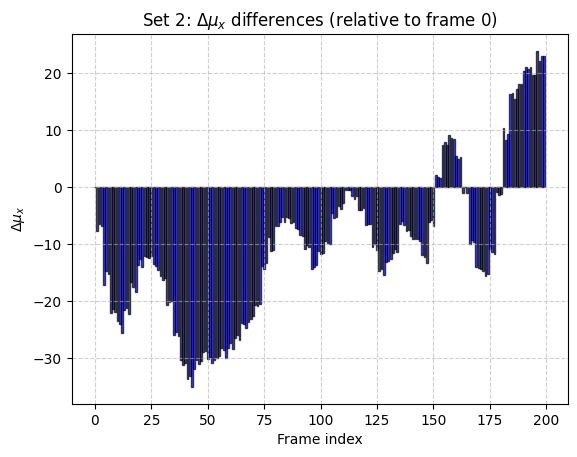}
            \caption{Relative Centroid Shift along 'X' Axis for Set 2}
        \end{subfigure}
        \hfill
        \begin{subfigure}[t]{0.30\textwidth}
            \centering
            \includegraphics[width=\linewidth]{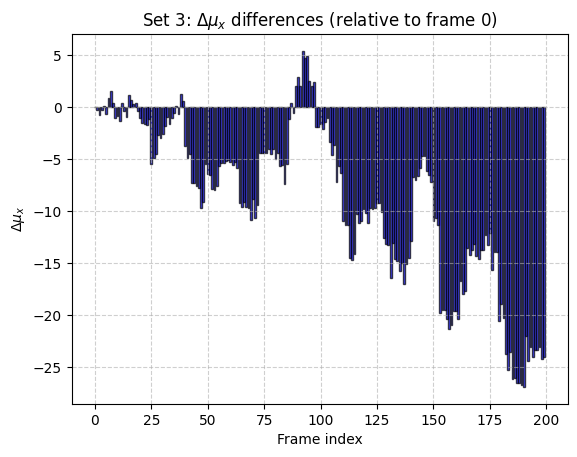}
            \caption{Relative Centroid Shift along 'X' Axis for Set 3}
        \end{subfigure}
        \caption{Relative Centroids shifts for three cases along 'X' axis. Set 1: Raw Turbulence, Set 2: Turbulence with 1 PMMA Rod, Set 3: Turbulence with 2 PMMA Rod.}
        \label{Cent3}
    \end{figure}
\begin{figure}[H]
        \centering
        \begin{subfigure}[t]{0.30\textwidth}
            \centering
            \includegraphics[width=\linewidth]{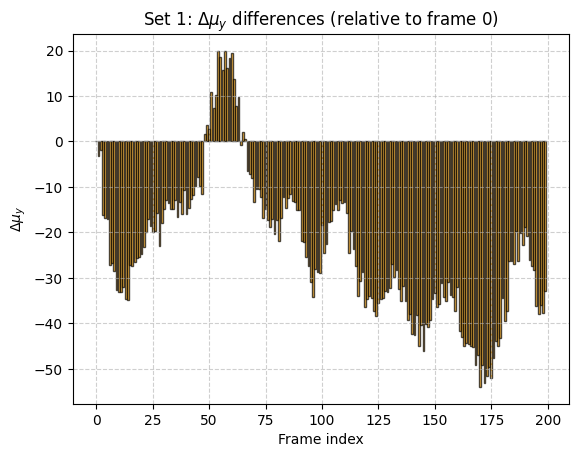}
            \caption{Relative Centroid Shift along 'Y' Axis for Set 1}
        \end{subfigure}
        \hfill
        \begin{subfigure}[t]{0.30\textwidth}
            \centering
            \includegraphics[width=\linewidth]{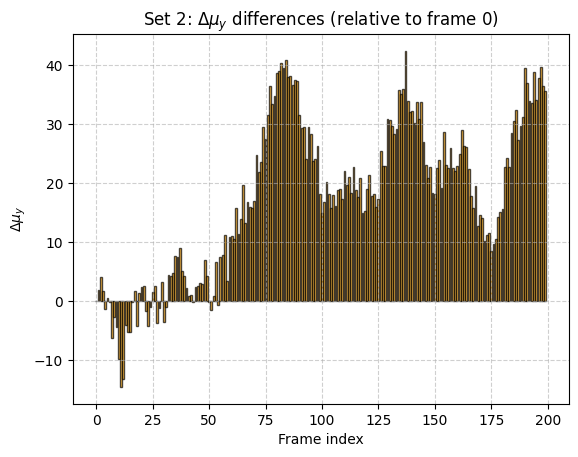}
            \caption{Relative Centroid Shift along 'Y' Axis for Set 2}
        \end{subfigure}
        \hfill
        \begin{subfigure}[t]{0.30\textwidth}
            \centering
            \includegraphics[width=\linewidth]{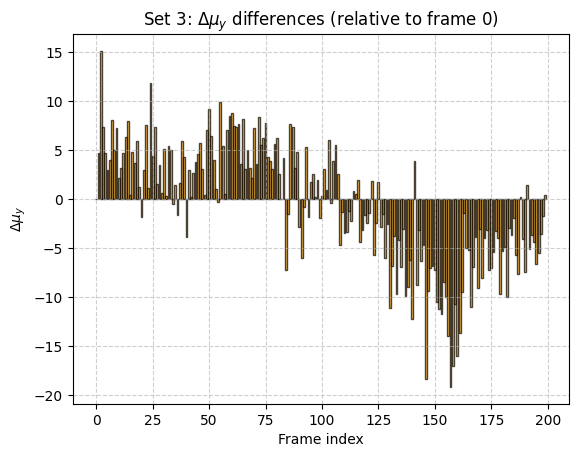}
            \caption{Relative Centroid Shift along 'Y' Axis for Set 3}
        \end{subfigure}
        \caption{Relative Centroids shifts for three cases along 'Y' axis. Set 1: Raw Turbulence, Set 2: Turbulence with 1 PMMA Rod, Set 3: Turbulence with 2 PMMA Rod.}
        \label{Cent4}
    \end{figure}

In the baseline case of raw turbulence, the centroid trajectories show large, irregular fluctuations. The absolute centroid positions $(\mu_x,\mu_y)$ wander significantly over the 200 frames, and the relative shifts $(\Delta \mu_x,\Delta \mu_y)$ demonstrate systematic drifts. Specifically, $\Delta \mu_x$ reaches values up to $+50$ units, while $\Delta \mu_y$ drifts downward to nearly $-50$ units. This indicates that in the absence of compensation, turbulence introduces strong low-frequency fluctuations as well as high-amplitude jitter, destabilizing the beam profile. When a single PMMA rod is introduced, the centroid dynamics are altered, but stability is not significantly improved. The plots of $(\mu_x,\mu_y)$ continue to show large excursions, and the relative shift $\Delta \mu_y$ exhibits a pronounced positive drift exceeding $+40$ units. Thus, while the presence of one dielectric rod modifies the turbulence-induced distortions, it does not suppress them effectively. The overall magnitude of centroid fluctuations remains comparable to the raw turbulence case, suggesting that a single PMMA rod is insufficient as a compensating mechanism. A marked improvement is observed when two PMMA rods are employed in sequence. In this case, the centroid positions $(\mu_x,\mu_y)$ remain tightly clustered around their initial values. The relative shifts $(\Delta \mu_x,\Delta \mu_y)$ are confined within a range of approximately $\pm 10$ units over the entire 200-frame sequence. This represents a significant reduction in both short-term jitter and long-term drift. The two-rod configuration therefore, demonstrates effective compensation against turbulence, stabilizing the centroid to near-baseline levels (Figures \ref{Cent3} and \ref{Cent4}).\par
The centroid stability analysis reveals a clear trend: raw turbulence induces severe instability, a single PMMA rod does not provide adequate correction, while the two-rod configuration yields substantial stabilization. These findings establish the two-rod PMMA setup as the most effective method for mitigating turbulence-induced beam centroid fluctuations. The presence of inertia forces opposes the frequent changes of dipole distributions inside the PMMA rods which finally decrease the fluctuations of relative centroid shifts in each case of experimental studies. These experimental outcomes produces direct proof of presence of dipole-dipole coupling dependent inertia forces and dipole oscillation synchronization inside the medium after propagation of light beam with pseudo-randomly spatial polarization distribution. 

\subsection{Variance \& Aberrated Wavefronts}
A detailed evaluation of the variance and standard deviation plots was conducted to assess how turbulence and dielectric compensation affect the spatial size and stability of the beam feature. The standard deviations $\sigma_x$ and $\sigma_y$ characterize the spread of the intensity distribution along the $x$- and $y$-axes, respectively. Relative variations were further quantified as $\Delta \sigma_x = \sigma_x - \sigma_x^{(0)}$ and $\Delta \sigma_y = \sigma_y - \sigma_y^{(0)}$, where $\sigma^{(0)}$ denotes the initial-frame values. The comparative analysis was performed across three experimental conditions: raw turbulence (Set~1), turbulence with one PMMA rod (Set~2), and turbulence with two PMMA rods (Set~3).\par
\begin{figure}[H]
        \centering
        \begin{subfigure}[t]{0.48\textwidth}
            \centering
            \includegraphics[width=\linewidth]{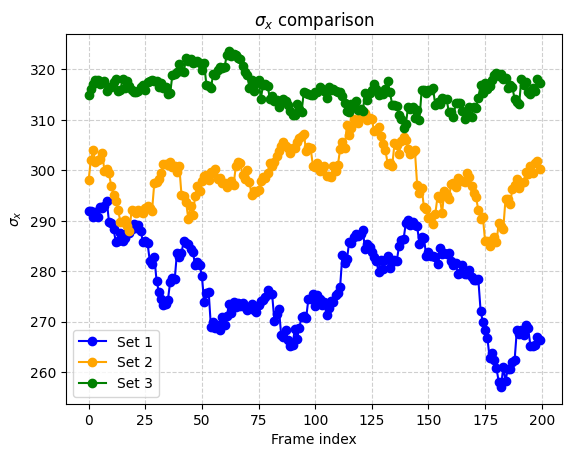}
            \caption{$\sigma_x$ from diagonal of Covariance Matrix}
        \end{subfigure}
        \hfill
        \begin{subfigure}[t]{0.48\textwidth}
            \centering
            \includegraphics[width=\linewidth]{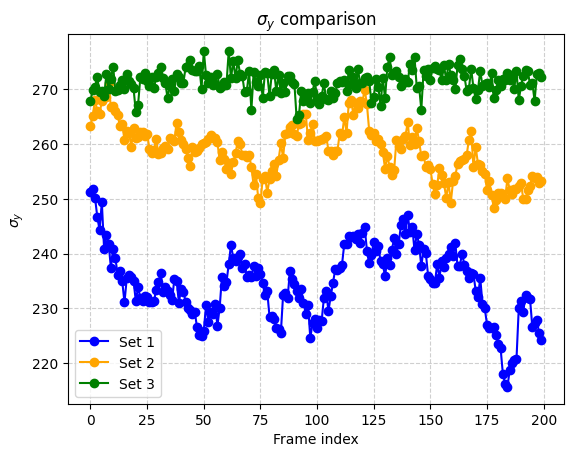}
            \caption{$\sigma_y$ from diagonal of Covariance Matrix}
        \end{subfigure}
        \caption{Standard Deviations for three cases. Set 1: Raw Turbulence, Set 2: Turbulence with 1 PMMA Rod, Set 3: Turbulence with 2 PMMA Rod.}
        \label{Var1}
    \end{figure}
\begin{figure}[H]
\centering
\begin{minipage}[b]{0.65\textwidth}
    \includegraphics[width=\textwidth]{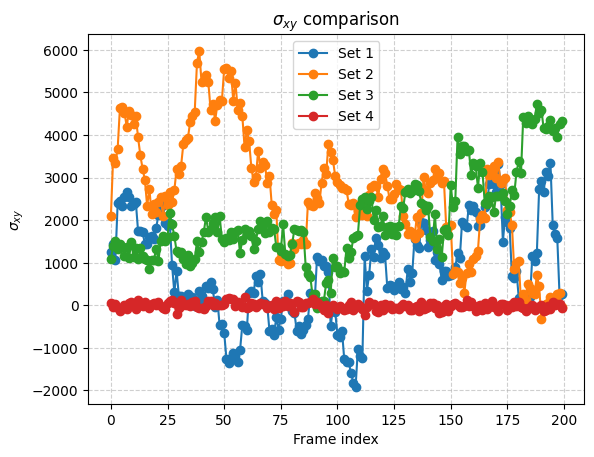}
    \caption{Off-diagonal Standard Deviations $(\sigma_{xy})$}
    \label{Var5}
\end{minipage}
\end{figure}

In the baseline condition with uncorrected turbulence, the beam feature exhibits strong instability and a clear shrinking tendency (figure: \ref{Var1}). The standard deviation values steadily drift downward, and the relative change plots show persistent negative trends. Both $\Delta \sigma_x$ and $\Delta \sigma_y$ decrease by more than $30$ units, with shrinkage reaching up to $-35$ units in some frames. This indicates that turbulence causes the beam feature to contract significantly over time, reducing its spatial extent and thereby destabilizing the propagation profile. When a single PMMA rod is inserted (figure: \ref{Var1}), the shrinking effect is replaced by large fluctuations, but stability is not achieved. The relative changes $\Delta \sigma_x$ and $\Delta \sigma_y$ oscillate erratically, spanning wide ranges. Specifically, $\Delta \sigma_x$ varies over approximately $24$ units (from $-12$ to $+12$), while $\Delta \sigma_y$ fluctuates across a $23$-unit interval (from $-15$ to $+8$). Thus, although the one-rod configuration prevents monotonic contraction, it introduces bidirectional instability that fails to regulate the feature size effectively. The configuration with two PMMA rods demonstrates both the largest feature size and the greatest stability (figure: \ref{Var1}). The relative changes are confined within narrow ranges, with $\Delta \sigma_x$ remaining between $-5$ and $+9$ (a $14$-unit spread) and $\Delta \sigma_y$ restricted to a tight interval of $0$ to $+10$ units. These results highlight the ability of the two-rod setup to suppress turbulent distortions, maintaining a consistently expanded beam profile with minimal fluctuations. The figure: \ref{Var5} presents the asymmetric counterpart of the covariance matrix. It presents non-zero covariance for turbulence impacted beams whereas zero covariance for the beam without turbulence impact.\par
The comparative results establish a clear trend: raw turbulence induces shrinking, a single PMMA rod produces erratic fluctuations, while two rods yield a larger and more stable feature size. Thus, the two-rod configuration provides the most effective compensation against turbulence-induced size instabilities. The increase in PMMA rods with non-polished output surface produces additional aberrations on the beam wavefronts. Hence, the presence of increasing aberrations with increase of number of PMMA rods increases the spreading of the beam wavefronts as well as intensity distributions. Thus, increase in spreading and aberration increases the weight of skewness and excess in the fitted curves. Hence, these results are direct proof of the phenomena that the output beam wavefront aberration is totally dependent upon medium output surface which again dependent upon the dipole moment and corresponding free space radiations from the output surface.

% Variance

\begin{figure}[H]
        \centering
        \begin{subfigure}[t]{0.48\textwidth}
            \centering
            \includegraphics[width=\linewidth]{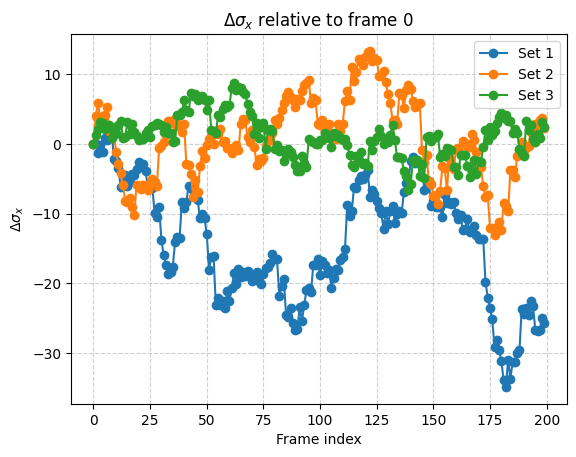}
            \caption{Relative Change in $\sigma_x$}
        \end{subfigure}
        \hfill
        \begin{subfigure}[t]{0.48\textwidth}
            \centering
            \includegraphics[width=\linewidth]{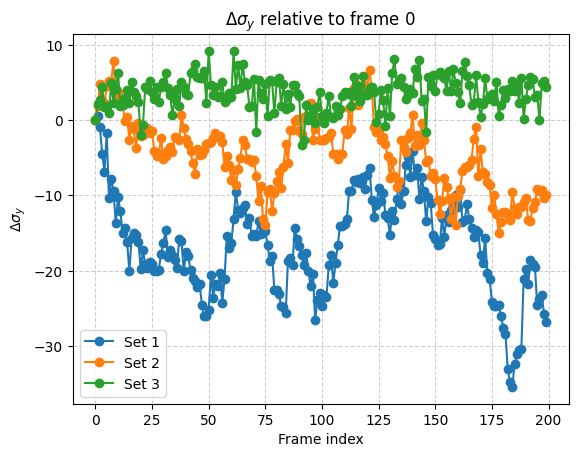}
            \caption{Relative Change in $\sigma_y$}
        \end{subfigure}
        \caption{Standard Deviations Relative Changes for three cases. Set 1: Raw Turbulence, Set 2: Turbulence with 1 PMMA Rod, Set 3: Turbulence with 2 PMMA Rod.}
        \label{Var2}
    \end{figure}
\begin{figure}[H]
        \centering
        \begin{subfigure}[t]{0.30\textwidth}
            \centering
            \includegraphics[width=\linewidth]{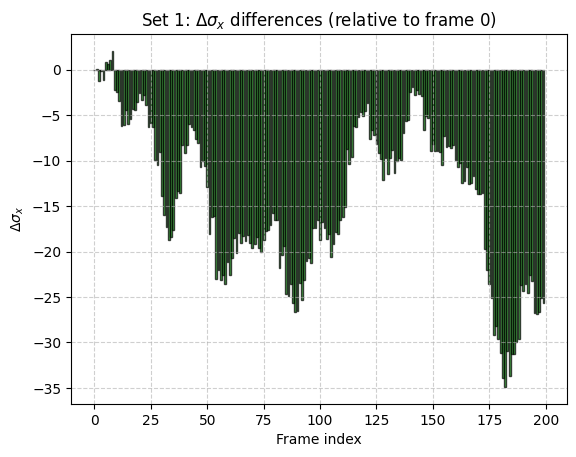}
            \caption{Relative Change in $\sigma_x$ for Set 1}
        \end{subfigure}
        \hfill
        \begin{subfigure}[t]{0.30\textwidth}
            \centering
            \includegraphics[width=\linewidth]{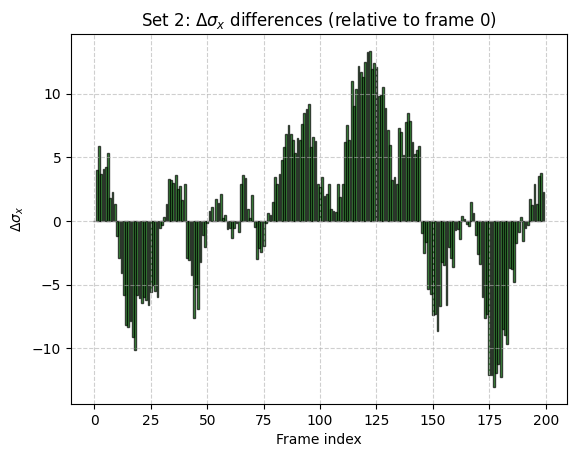}
            \caption{Relative Change in $\sigma_x$ for Set 2}
        \end{subfigure}
        \hfill
        \begin{subfigure}[t]{0.30\textwidth}
            \centering
            \includegraphics[width=\linewidth]{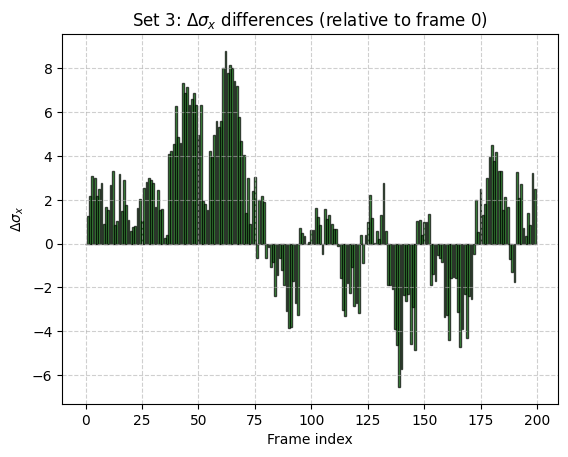}
            \caption{Relative Change in $\sigma_x$ for Set 3}
        \end{subfigure}
        \caption{Standard Deviations Relative Changes for three cases. Set 1: Raw Turbulence, Set 2: Turbulence with 1 PMMA Rod, Set 3: Turbulence with 2 PMMA Rod.}
        \label{Var3}
    \end{figure}
\begin{figure}[H]
        \centering
        \begin{subfigure}[t]{0.30\textwidth}
            \centering
            \includegraphics[width=\linewidth]{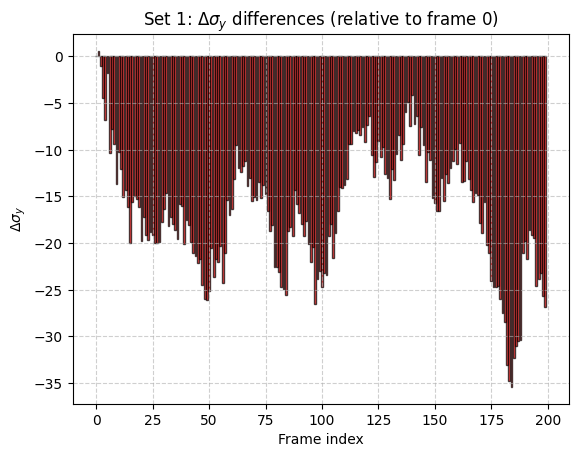}
            \caption{Relative Change in $\sigma_y$ for Set 1}
        \end{subfigure}
        \hfill
        \begin{subfigure}[t]{0.30\textwidth}
            \centering
            \includegraphics[width=\linewidth]{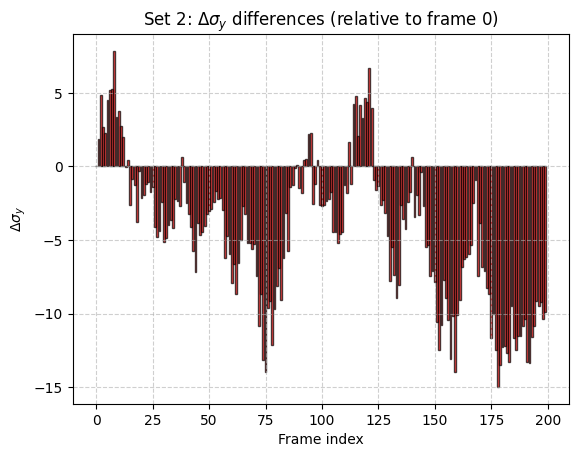}
            \caption{Relative Change in $\sigma_y$ for Set 2}
        \end{subfigure}
        \hfill
        \begin{subfigure}[t]{0.30\textwidth}
            \centering
            \includegraphics[width=\linewidth]{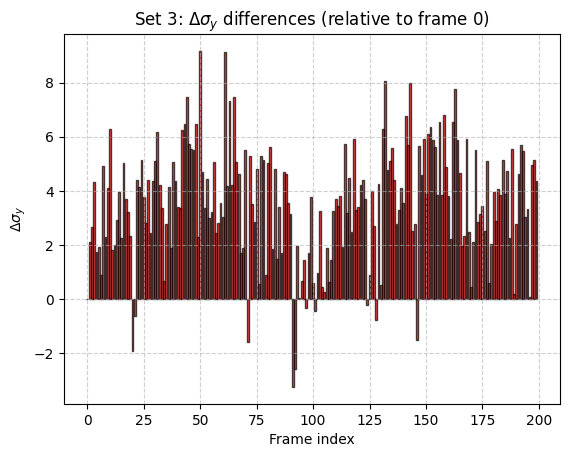}
            \caption{Relative Change in $\sigma_y$ for Set 3}
        \end{subfigure}
        \caption{Standard Deviations Relative Changes for three cases. Set 1: Raw Turbulence, Set 2: Turbulence with 1 PMMA Rod, Set 3: Turbulence with 2 PMMA Rod.}
        \label{Var4}
    \end{figure}

\subsection{Power \& Volume Scintillation}
The results obtained from the volume (signal power) and Scintillation Index (SI) plots provide complementary perspectives on the stability of the optical system under different turbulence compensation schemes. The volume, defined as the integrated intensity under the fitted Gaussian distribution, is proportional to the total power transmitted through the system. In contrast, the Scintillation Index quantifies the normalized variance of intensity fluctuations, serving as a direct indicator of turbulence strength. Together, these two measures demonstrate that the two-PMMA-rod configuration (Set~3) is the most effective at mitigating turbulence effects.\par
\begin{figure}[H]
\centering
\begin{minipage}[b]{0.75\textwidth}
    \includegraphics[width=\textwidth]{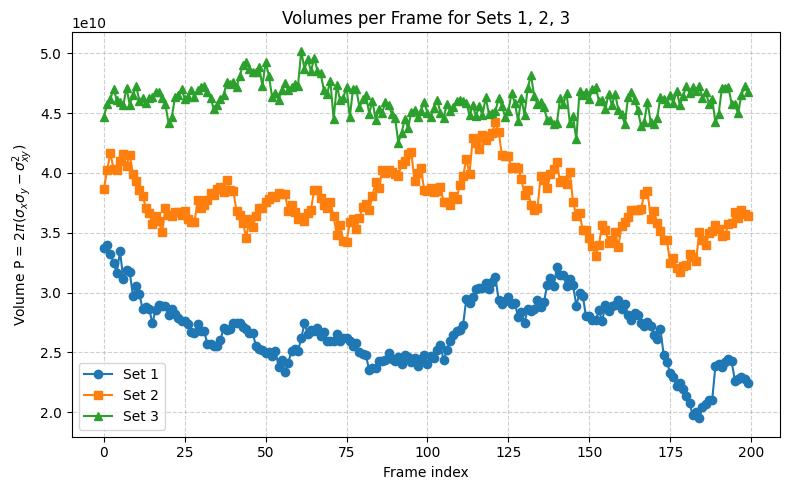}
    \caption{Volume under the fitted Functions for all frames for three cases. Set 1: Raw Turbulence, Set 2: Turbulence with 1 PMMA Rod, Set 3: Turbulence with 2 PMMA Rod $(|\Sigma|=\sigma_x^2\sigma_y^2-\sigma^2_{xy})$. Magnitude order $10^5$}
    \label{Pow1}
\end{minipage}
\end{figure}
\begin{figure}[H]
\centering
\begin{minipage}[b]{0.75\textwidth}
    \includegraphics[width=\textwidth]{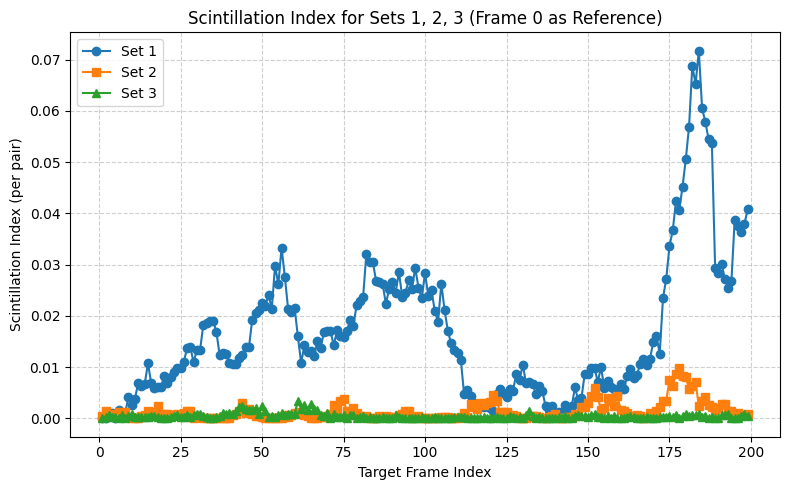}
    \caption{Relative Scintillation Index between pairs for three cases. Set 1: Raw Turbulence, Set 2: Turbulence with 1 PMMA Rod, Set 3: Turbulence with 2 PMMA Rod. Each pairs contains a reference frame (0th Frame) and a target frame (all other frames).}
    \label{Pow2}
\end{minipage}
\end{figure}

The analysis of total power reveals clear distinctions among the three experimental conditions (Figure: \ref{Pow1}). In Set~3 (two rods), the system maintains the highest and most stable power levels across the 200 recorded frames. The absence of a significant downward trend indicates that this configuration effectively suppresses energy dissipation and signal degradation caused by turbulence. In contrast, Set~1 (raw turbulence) displays the lowest average power accompanied by large fluctuations. A gradual decline in signal strength is evident, confirming that turbulence without compensation leads to both substantial and unstable energy loss. Set~2 (one rod) exhibits intermediate performance: while some improvement is observed compared to the raw turbulence case, the power profile remains noticeably less stable and less efficient than that achieved with two rods. Higher SI values correspond to stronger turbulence-induced fluctuations. In Set~1, the SI reaches highly erratic levels, with peaks exceeding $0.07$, reflecting strong and uncontrolled turbulence. The introduction of one PMMA rod (Set~2) markedly reduces the scintillation, lowering the SI below $0.01$ and demonstrating a partial suppression of turbulence. The most notable result is obtained with Set~3 (two rods), where the SI remains consistently near zero throughout the sequence. This signifies an almost complete elimination of turbulence-induced intensity fluctuations and a high degree of signal stability (Figure: \ref{Pow2}). In summary, both the power preservation results and the Scintillation Index analysis converge to the same conclusion: the two-rod PMMA configuration offers superior turbulence mitigation, ensuring a high-power, highly stable output signal.

% Power


\section{Conclusion}\label{6}
This work has successfully developed a comprehensive theoretical and experimental framework to demonstrate the mitigation of turbulence-induced scintillation in laser beams through collective dipole oscillations in a dielectric medium. The theoretical foundation, built upon an extended Lorentz oscillator model incorporating nonlinear restoring forces and dipole-dipole interactions via the dyadic Green's function, predicted that synchronized dipole modes within a dense medium could impose an inertial effect, thereby stabilizing a propagating optical field against turbulent fluctuations. Experimental validation was conducted using a pseudo-random phase plate to simulate turbulence and PMMA rods as the dielectric medium. The results unequivocally support the theoretical predictions. The baseline case with raw turbulence exhibited severe degradation, characterized by large centroid wander (up to 50 units), significant power loss, and a high Scintillation Index (peaking above 0.07). The introduction of a single PMMA rod proved insufficient for effective compensation. However, the configuration with two PMMA rods demonstrated remarkable performance. This setup effectively suppressed beam wander, confining centroid shifts to a minimal range of approximately ±10 units. It also counteracted the beam shrinking observed in the raw turbulence case, maintaining a larger and more stable spatial profile. Most significantly, the two-rod system preserved the highest signal power and reduced the Scintillation Index to a near-zero value, indicating an almost complete cancellation of intensity fluctuations. In conclusion, this study validates that leveraging the collective, synchronized dynamics of dipoles in a dielectric material offers a robust and passive method for scintillation compensation. The strong agreement between the advanced theoretical model and the experimental outcomes establishes this phenomenon as a promising new avenue for enhancing the reliability and performance of free-space optical communication systems operating in turbulent environments.

\section*{Funding}
Department of Science and Technology, Ministry of Science and Technology, India (CRG/2020/003338).

\section*{Declaration of competing interest}
The authors declare the following financial interests/personal relationships which may be considered as potential competing interests: Shouvik Sadhukhan reports a relationship with Indian Institute of Space Science and Technology that includes: employment. NA If there are other authors, they declare that they have no known competing financial interests or personal relationships that could have appeared to influence the work reported in this paper.

\section*{Data availability}
All data used for this research has been provided in the manuscript itself.

\section*{Acknowledgments}
Shouvik Sadhukhan and C S Narayanamurthy Acknowledge the SERB/DST (Govt. Of India) for providing financial support via the project grant CRG/2020/003338 to carry out this work. Shouvik Sadhukhan would like to thank Mr. Amit Vishwakarma and Dr. Subrahamanian K S Moosath from Department of Mathematics, Indian Institute of Space Science and Technology Thiruvananthapuram for their suggestions into statistical analysis in this paper.

\section*{CRediT authorship contribution statement}
\textbf{Shouvik Sadhukhan:} Writing– original draft, Visualization, Formal analysis. \textbf{C. S. Narayanamurthy:} Writing– review $\&$ editing, Validation, Supervision, Resources, Project administration, Investigation, Funding acquisition, Conceptualization.

%\section*{Appendix I}

\end{document}